\documentclass[pra,english,preprintnumbers,nofootinbib,twocolumn,superscriptaddress]{revtex4-2}
\usepackage[utf8]{inputenc}
\usepackage[T1]{fontenc}
\usepackage{amsmath}
\usepackage{amssymb}
\usepackage{graphicx}
\usepackage{tabularx}
\usepackage{bm}
\usepackage{pifont}
\usepackage{xifthen}
\usepackage{placeins}
\usepackage{xcolor}
\usepackage[colorlinks=true,linkcolor=blue]{hyperref}

\newcommand{\kf}[1][]{k_{{\mathrm F}\ifthenelse{\equal{#1}{}}{}{,#1}}}
\newcommand{\tf}[1][]{T_{{\mathrm F}\ifthenelse{\equal{#1}{}}{}{,#1}}}
\newcommand{\ef}[1][]{\varepsilon_{{\mathrm F}\ifthenelse{\equal{#1}{}}{}{,#1}}}
\newcommand{\efg}[1][]{E_{{\mathrm FG}\ifthenelse{\equal{#1}{}}{}{,#1}}}

\newcommand{\up}{\uparrow}
\newcommand{\down}{\downarrow}
\newcommand{\etal}{{\it et al.}}
\newcommand{\tr}{\mathrm{Tr}}

\newcommand{\f}[2]{\frac{#1}{#2}}
\newcommand{\ev}[1]{\langle#1\rangle}
\newcommand{\pf}[2]{\frac{\partial #1}{\partial #2}}
\newcommand{\tpf}[3]{\left( \pf{#1}{#2} \right)_{#3}}

\newcommand{\mZ}{\mathcal{Z}}
\newcommand{\mD}{\mathcal{D}}
\newcommand{\e}{{\rm e}}
\renewcommand{\d}{{\rm d}}

\newcommand{\equref}[1]{Eq.~(\ref{eq:#1})}
\newcommand{\figref}[1]{Fig.~\ref{fig:#1}}
\newcommand{\tabref}[1]{Tab.~\ref{tab:#1}}
\newcommand{\appref}[1]{App.~\ref{app:#1}}

\newcommand{\beq}{\begin{equation}}
\newcommand{\eeq}{\end{equation}}
\newcommand{\bea}{\begin{eqnarray}}
\newcommand{\eea}{\end{eqnarray}}

\usepackage{booktabs}
\usepackage{array}
\usepackage{arydshln}

\newcommand{\PreserveBackslash}[1]{\let\temp=\\#1\let\\=\temp}
\newcolumntype{C}[1]{>{\PreserveBackslash\centering}p{#1}}
\newcolumntype{M}[1]{>{\centering\arraybackslash}m{#1}}
\newcolumntype{L}[1]{>{\arraybackslash}m{#1}}
\newcolumntype{N}{@{}m{0pt}@{}}

\hypersetup{
	colorlinks,
	linkcolor={blue!75!black},
	citecolor={blue!75!black},
	urlcolor={blue!75!black},
}

\begin{document}

\title{Pairing and the spin susceptibility of the polarized \\ unitary Fermi gas in the normal phase}

\author{Lukas Rammelm\"uller}
\affiliation{Arnold Sommerfeld Center for Theoretical Physics (ASC), University of Munich, Theresienstr. 37, 80333 M\"unchen, Germany}
\affiliation{Munich Center for Quantum Science and Technology (MCQST), Schellingstr. 4, 80799 M\"unchen, Germany}

\author{Yaqi Hou}
\affiliation{Department of Physics and Astronomy, University of North Carolina, Chapel Hill, North Carolina 27599, USA}

\author{Joaqu\'in E. Drut}
\affiliation{Department of Physics and Astronomy, University of North Carolina, Chapel Hill, North Carolina 27599, USA}

\author{Jens Braun}
\affiliation{Institut f\"ur Kernphysik (Theoriezentrum), Technische Universit\"at Darmstadt, D-64289 Darmstadt, Germany}
\affiliation{FAIR, Facility for Antiproton and Ion Research in Europe GmbH, Planckstra{\ss}e 1, D-64291 Darmstadt, Germany}
\affiliation{ExtreMe Matter Institute EMMI, GSI, Planckstra{\ss}e 1, D-64291 Darmstadt, Germany}

\begin{abstract}
We theoretically study the pairing behavior of the unitary Fermi gas in the normal phase. Our analysis is based on the static spin susceptibility, which characterizes the response to an external magnetic field. We obtain this quantity by means of the complex Langevin approach and compare our calculations to available literature data in the spin-balanced case. Furthermore, we present results for polarized systems, where we complement and expand our analysis at high temperature with high-order virial expansion results. The implications of our findings for the phase diagram of the spin-polarized unitary Fermi gas are discussed, in the context of the state of the art.
\end{abstract}

\maketitle

\section{Introduction}
Pair formation in fermionic quantum matter is at the heart of a wide range of physical phenomena and pervades physics at vastly different lengths scales, ranging from ultracold quantum gases to neutron stars. In \mbox{spin-$\tfrac{1}{2}$} systems with weak attractive interactions, for example, pairing of spin-up and spin-down particles at diametrically opposite points of the Fermi surface (i.e., Cooper pairing) typically leads to a superfluid phase at low enough temperatures, accompanied by the opening of an energy gap in the quasiparticle spectrum.
Similar conclusions hold for strongly correlated systems, where pairing and superfluidity have been addressed thoroughly in theory~\cite{Perali2002,Chen2005,Bulgac2006,Stewart2008,Kuhnle2010,Perali2011,Palestini2012,Enss2012b,Boettcher2013,Wlazlowski2013,Tajima2014,Pantel2014,Chen2014,Boettcher2014,Roscher2015,Krippa2015,Jensen2020,Jensen2019,Pini2019,Richie-Halford2020} and experiment~\cite{Schunck2008,Shin2008,Nascimbene2010,Nascimbene2010b,Sommer2011,Nascimbene2011,Ku2012,Sagi2015,Horikoshi2017,Murthy2018,Jensen2019b,Long2020}. A central open question in this regard is whether signatures of strong pairing fluctuations survive in the normal, non-superfluid high-temperature phase, which is often referred to as a ``pseudogap regime'', obtaining its name from a (potentially strong) suppression of the single-particle density of states around the Fermi surface.

An intriguing system in the above sense is the unitary Fermi gas (UFG), a spin-$\tfrac{1}{2}$ system with a zero-range attractive interaction tuned to the threshold of two-body bound-state formation~\cite{Zwerger2012}. This strongly correlated system features non-relativistic conformal invariance~\cite{Nishida2007} and universal properties~\cite{Ho2004,Braaten2006} that have attracted researchers from several areas across physics.
The thermodynamics of this system has been investigated via a broad range of theoretical methods~\cite{Zwerger2012} and many of its properties have been measured experimentally (see e.g.~\cite{ExpReview}). However, a number of open questions remain, in particular regarding the appearance or not of the above-mentioned pseudogap~\cite{Gaebler2010, Sagi2015}.

There is consensus on the qualitative features of a ``pseudogap phase'', which is expected to leave an imprint on quantities such as odd-even staggering of the energy per particle and suppression of the spin susceptibility above the critical temperature $T_c$.
However, there is no long-range order parameter associated with the emergence of a such a behavior. An unambiguous definition of this ``phase'' therefore does not exist. The case of the UFG is especially challenging in this regard, as it is in a strongly correlated non-ordered (i.e., non-superfluid) phase above the critical temperature $T_c$ (see Ref.~\cite{Mueller2017} for a discussion and Refs.~\cite{Chen2014,Jensen2019} for recent reviews). Moreover, as argued in Ref.~\cite{Rothstein2019}, the UFG is not described by (normal) Fermi liquid theory at temperatures between $T_c$
and the Fermi temperature $\tf$.

Consequently, the debate remains active due to the varying opinions as to what the magnitude of the various signals for a ``pseudogap phase'' should be (e.g., how much suppression in the density of states or the susceptibility truly constitutes a pseudogap). This issue is further complicated by the lack of a small expansion parameter or effective theory that would allow to study this regime in a systematic manner.
Thus, to face this challenge and shed light on the pairing properties of the UFG in the normal phase, one must use nonperturbative methods that start from the microscopic degrees of freedom, i.e., {\it ab initio} approaches.

In the superfluid, low-temperature phase, pairing is loosely speaking parameterized by the pairing gap. However, at temperatures above the phase transition, the gap is necessarily zero in the long-range limit and therefore this quantity does not allow us to quantify pairing effects unambiguously. Several quantities have instead been proposed for that purpose.
A prominent example is the spectral function $A({\bf k},\omega)$ which should be suppressed around the Fermi energy in the presence of extensive pairing fluctuations. While, e.g., T-matrix studies report a substantial suppression of spectral weight~\cite{Perali2002,Pieri2004}, self-consistent Luttinger-Ward (LW) studies have found almost no signature of a pseudogap in this quantity~\cite{Haussmann2009}. Notably, the density of states $\rho(\omega) = \int\tfrac{\d^3 k}{(2\pi)^3}\ A({\bf k},\omega)$ was also computed via an auxiliary-field quantum Monte Carlo (AFQMC) approach combined with a numerical analytic continuation to real frequencies~\cite{Wlazlowski2013}. There, a distinct suppression was found. However, those results were obtained with a spherical cutoff in momentum space which could lead to unphysical results (see Ref.~\cite{Jensen2019,Jensen2020} for a discussion).

As an alternative probe to study pairing above the superfluid transition, one may focus on thermodynamic response functions, such as the magnetic (or spin) susceptibility~$\chi$:
\begin{equation}
  \chi \equiv \tpf{M}{h}{T,V,\mu} = \beta\left(\ev{M^2} - \ev{M}^2\right)\,,
  \label{eq:chi_m}
\end{equation}
where $M = N_\up - N_\down$ denotes the magnetization of the system. In the context of cuprate superconductors, it was suggested that the pairing gap would suppress $\chi$, making it a good candidate to study pairing in the normal phase~\cite{Trivedi1995,Randeria1992}. Similar conclusions hold for the specific heat, see, e.g., Ref.~\cite{Jensen2019,Jensen2020}.

The spin susceptibility allows for a clear definition of a spin-gap temperature $T_s$ based on the following arguments. The spin susceptibility vanishes as $T\to0$ as a consequence of the presence of a pairing gap in the quasiparticle spectrum of a fermionic superfluid (or superconductor).
At high temperatures, on the other hand, the system enters the Boltzmann regime and decays following Curie's law $\chi = C T^{-1}$, where $C$ is the material-specific Curie constant. Therefore, $\chi$ necessarily develops a maximum at some intermediate temperature, which defines the aforementioned spin-gap temperature $T_s$ (see Ref.~\cite{Tajima2014} for a discussion).
Note that the suppression of $\chi$ in the superfluid phase was dubbed ``anomalous'' since the noninteracting case approaches the Pauli susceptibility $\chi_{\mathrm{P}} = \f{3}{2}\f{n}{\ef}$ monotonically as $T\to 0$ (where $n$ and $\ef$ are the density and Fermi energy of the noninteracting case, respectively). The spin-gap temperature $T_s$ may be interpreted as the highest temperature at which pairing correlations may strongly influence the dynamics of the system. For~$T>T_s$, the spin susceptibility is dominated by thermal fluctuations and approaches the noninteracting limit.

In this work, we use the complex Langevin~(CL) approach~\cite{Berger2021,Attanasio2020} to characterize the spin susceptibility and spin-gap temperature of the polarized UFG and gain insight into pair formation above $T_c$. Previous non-perturbative ab initio investigations~\cite{Jensen2020,Richie-Halford2020} have also calculated the susceptibility as well as odd-even staggering and the single-particle spectrum, but focused on the unpolarized case. Here, we review and comprehensively compare extant susceptibility data for the unpolarized system and extend the presentation to the polarized case, showing results for several temperatures and polarizations. In addition, we validate our findings at high temperature using a Pad\'e-resummed virial expansion (PRVE) analysis~\cite{Hou2020b, Hou2020}. Finally, we set our findings in context by examining implications for the phase-diagram of the spin-polarized Fermi gas.
%

\section{Model}
We consider a spin-$\tfrac{1}{2}$ Fermi gas with a short-range attractive interaction between spin-up and spin-down particles. The Hamiltonian describing this system (in natural units $\hbar = k_{\rm B} = m = 1$) reads
\begin{eqnarray}
  \hat H &=&  -\frac{1}{2}\sum_{s=\uparrow,\downarrow}\! {\int{\!d^3 x\,\hat{\psi}^{\dagger}_{s}({\bf x})\nabla^2\hat{\psi}^{}_{s}({\bf x})}}\nonumber \\
 && \qquad\qquad\qquad - g\int{\!d^3 x\,\hat{n}^{}_{\uparrow}({\bf x})\hat{n}^{}_{\downarrow}({\bf x})}\,,
 \label{eq:Hdef}
\end{eqnarray}
where~$g$ is the bare coupling, $\hat{\psi}^{\dagger}_{s}({\bf x})$ and $\hat{\psi}_{s}({\bf x})$ create and annihilate fermions of spin $s$ at position ${\bf x}$, respectively, and $\hat n_s({\bf x}) = \hat{\psi}^{\dagger}_{s}({\bf x})\hat{\psi}^{}_{s}({\bf x})$ is the density operator.

Formally, this model is a simple non-relativistic interacting field theory which has been widely studied and therefore many of its gross features are well known. As a function of the coupling strength, in particular, this model realizes the so-called BCS-BEC crossover~\cite{Zwerger2012,Strinati2018}, given by a smooth evolution from a gas of weakly attractive fermions (BCS regime) to a gas of weakly repulsive composite bosons (BEC regime), typically parameterized by the dimensionless parameter $(\kf a_s)^{-1}$. Here, $\kf$ denotes the Fermi momentum which is determined by the total particle density $n$ via $\kf = (3\pi^2 n)^{1/3}$ and $a_s$ denotes the $s$-wave scattering length. When $a_s$ diverges, the system is at the threshold of two-body bound-state formation and the scattering cross-section assumes its maximal value (only bounded by unitarity of the scattering matrix) implying strong interactions. Since $a_s$ drops out of the problem in that limit, the density and the temperature are the only physical parameters determining the physical behavior. The system is then said to exhibit \emph{universality} in the sense that the microscopic details of the interaction become irrelevant.

The leading instability of the model is toward a superfluid state below a (coupling-dependent) critical temperature $T_c$ throughout the entire BCS-BEC crossover. In the limit of weak attraction, mean-field theory is applicable and the situation is well described by BCS theory. In that regime, pair formation (i.e., the appearance of a pairing gap) and pair condensation (i.e., the appearance of off-diagonal long-range order and superfluidity) happen simultaneously, i.e., the so-called pairing temperature $T^*$ is identical to $T_c$. On the other side of the resonance associated with the limit of an infinite $s$-wave scattering length, the interatomic potential
supports a two-body bound state, such that the relevant degrees of freedom are bound $\up\down$-molecules that form at $T^* \gg T_c$ and condense below $T_c$ to form a superfluid. While the situation appears to be comparatively clear in the aforementioned two limits, the situation at unitarity above $T_c$ has
remained ambiguous.
In fact, the question of whether strong pairing correlations survive above $T_c$ is intrinsically challenging due to strong interactions in the unitary limit. Non-perturbative methods, either stochastic approaches or semi-analytic methods (including resummation techniques) are therefore ultimately required to analyze the UFG even above the phase transition.

\section{Many-body methods}
In this section we briefly introduce the two theoretical approaches that we use to analyze the spin susceptibility of the polarized unitary Fermi gas in the normal phase.
\subsection{Lattice field theory}
To evaluate the thermodynamics of the UFG in a nonperturbative fashion, we use the CL approach. The partition function is given by
\begin{equation}
 \mZ = \tr\left[\e^{-\beta(\hat H - \mu_\up \hat N_\up - \mu_\down \hat N_\down)}\right]\,,
\end{equation}
where $\beta=1/T$ is the inverse temperature. Moreover, we have introduced the chemical potential $\mu_s$ and the particle-number operator $\hat{N}_s$ for the spin projection $s$. With this as a starting point, we regularize our Hamiltonian by putting the problem on a spatial lattice of extent $V =N_x \times N_x\times N_x$, where we set the lattice spacing $\ell = 1$ for all directions. We also employ a Suzuki-Trotter decomposition to discretize the imaginary time direction into $N_\tau = \beta / \Delta\tau$ slices, leading to a $(3+1)$ dimensional spacetime lattice of extent $N_x^3\times N_\tau$. Crucially, we can now employ a Hubbard-Stratonovich transformation to rewrite the quartic interaction term in Eq.~\eqref{eq:Hdef} as a one-body operator coupled to a bosonic auxiliary field $\phi$. This allows us to integrate out the fermions and thus express the above partition function as
\begin{equation}
  \mZ = \int \mD \phi\ \det \left(\mathcal{M}_\up[\phi]\mathcal{M}_\down[\phi] \right) \equiv \int \mD \phi\ \e^{-S[\phi]},
\end{equation}
where $\mD\phi = \prod_{\tau,{\bf i}} d \phi_{\tau,{\bf i}}$ is the field integration measure. Note that the product runs over all space-time lattice points. This integral is amenable to Monte Carlo (MC) importance sampling free from the sign-problem when $\det \mathcal{M}_\up[\phi] = \det \mathcal{M}_\down[\phi]$, i.e., in the SU(2) symmetric case. In the case of unequal spin populations, which is the case for unequal chemical potentials in the grand canonical ensemble, we may resort to the CL approach. The latter is able to deliver accurate results despite the possibly severe sign problem, as recently demonstrated for a variety of non-relativistic fermionic systems~\cite{Loheac2017,Rammelmueller2017,Rammelmueller2018,Rammelmueller2020}. Further details on the lattice approach to ultracold fermions as well as on CL can be found in Refs.~\cite{Drut2013,Lee2017,Berger2021,Attanasio2020}.

To tune the system to the unitary point, i.e., the limit of infinite $s$-wave scattering length, we choose the bare coupling constant $g$ in Eq.~\eqref{eq:Hdef} such that the lowest eigenvalue of the lattice two-body problem matches the eigenvalue obtained from the exact solution in a continuous finite volume, which is given by L{\"u}scher's formula~\cite{Luescher1986a,Luescher1986b}.

In this work, we use periodic boxes of linear extent $N_x = 11$, which was previously found adequate for the study of the normal phase of the UFG~\cite{Rammelmueller2018}. The inverse temperature is set to $\beta = 8.0$, such that the thermal wavelength $\lambda_T = \sqrt{2\pi\beta} \approx 7.1$ falls in the appropriate window $\ell \ll \lambda_T \ll N_x$ for physics in the continuum limit. The temporal discretization was set to $\tau = 0.05\ell^2$. Note that this choice of lattice parameters was previously used in several studies and found to be suitable~\cite{Burovski2006,Bulgac2008,Lee2009,Drut2011,Rammelmueller2018}.
We fix the adaptive target discretization of the Langevin time to values in the range $\Delta t_{\rm L} = 0.04$ to $0.16$ and extrapolate (linearly) to the limit $\Delta t_{\rm L}\to 0$. In our implementation of CL, we also included a regulator term to suppress uncontrolled excursions of the solver into the complex plane. The strength of this regulator term can be quantified in terms of a parameter~$\xi$ which we set to~$\xi = 0.1$, see Refs.~\cite{Loheac2017,Rammelmueller2017,Rammelmueller2018,Rammelmueller2020,Berger2021} for a detailed discussion. With respect to the spin susceptibility, we found that the results are independent of the regulator within reasonable bounds. All results presented below reflect averages of approximately $4000$ decorrelated samples, which gives a statistical uncertainty on the order of~$1-2$\%.

\subsection{Virial expansion}
The dilute limit associated with the limit of vanishing fugacity $z = e^{\beta \mu} \to 0$ (with $\mu = (\mu_\uparrow + \mu_\downarrow)/2$) can be addressed using the virial expansion, where the ratio of the interacting partition function $\mZ$ to its noninteracting value $\mathcal Z_0$ is written as a series in powers of $z$ (see Ref.~\cite{Liu2013} for review). More specifically, for an unpolarized system we have
\bea
\label{Eq:ZZ0virial}
\ln(\mathcal{Z} / \mathcal{Z}_0 ) = Q_1 \sum_{n=1}^{\infty} \Delta b_n z^n,
\eea
Here, $\Delta b_n$ is the interaction-induced change in the \mbox{$n$-th} virial coefficient and $Q_1 = 2V/\lambda_T^3$ is the single-particle partition function. While $\Delta b_2$ has been known for arbitrary coupling strengths since the 1930's~\cite{Beth1937} (see also~\cite{Lee2006}), results for $\Delta b_3$ only became available in this century~\cite{Liu2009, Bedaque2003, Kaplan2011, Leyronas2011, Castin2013, Gao2015}, those for $\Delta b_4$ (only at unitarity) within the last decade~\cite{Rakshit2012,Endo2015,Yan2016,Endo2016,Ngampruetikorn2015}, and, more recently, calculations  from weak coupling to unitarity were extended up to $\Delta b_5$~\cite{Hou2020b, Hou2020}. At a given order $n$, $\Delta b_n$ accounts for the interaction effects of the $n$-particle subspace of the Fock space. For the spin-$1/2$ system of interest here, each $\Delta b_n$ can be further broken down into polarized contributions $\Delta b_{mj}$, with $m+j = n$, coming from the subspace with $m$ spin-up particles and $j$ spin-down particles. In this work, we use the \( \Delta b_n \) calculated in Ref.~\cite{Hou2020b}. We note that $\Delta b_{21}$ was calculated in Ref.~\cite{Liu2009}, and \( \Delta b_{31} \) and \( \Delta b_{22} \) in Ref.~\cite{Yan2016} (see also \cite{Liu2010,Endo2020}). These quantities are essential for our purpose here as they are needed to calculate the properties of polarized matter. Indeed, in this case, Eq.~(\ref{Eq:ZZ0virial}) becomes
\bea
\ln(\mathcal{Z} / \mathcal{Z}_0 ) = Q_1 \sum_{n=2}^{\infty} \sum_{\substack{m,j > 0 \\ m + j  = n}} \Delta b_{mj} z_{\uparrow}^m z_{\downarrow}^j\,,
\eea
where $z_{s} = e^{\beta \mu_s}$ and~$s=\uparrow, \downarrow$. Using the knowledge of $\Delta b_{mj}$ and differentiating with respect to $\beta \mu_s$, we obtain expressions for the density, polarization, compressibility, and spin susceptibility within the virial expansion. In particular, the interaction effect on the individual spin particle numbers is given by
\bea
\Delta N_s &=& Q_1\sum_{n=2}^{\infty} \sum_{\substack{m,j > 0 \\ m + j  = n}}  (m \delta_{s\uparrow} + j \delta_{s\downarrow} ) \, \Delta b_{mj} z_{\uparrow}^m z_{\downarrow}^j\, ,
\eea
where \( \Delta N_s = \partial \ln(\mathcal{Z} / \mathcal{Z}_0) /  \partial (\beta \mu_s) \).
From this, one obtains the particle number of the species associated with spin~$s$ as $N_s = N_0(z_s) + \Delta N_s$, where $N_0(z_s)$ is the particle number for a noninteracting spinless system at fugacity $z_s$. Using the latter expressions, the total particle number in the interacting system is $N = N_\uparrow + N_\downarrow$, the magnetization is $M = N_\uparrow - N_\downarrow$, and the relative polarization is $p = M/N$. In the same way, the interaction effect on the susceptibility is given by
\bea
\begin{aligned}
\Delta \chi = \frac{\lambda_T^2}{8 \pi} Q_1 \sum_{n=3}^{\infty} \sum_{m+j=n} (m - j)^2 \Delta b_{mj} z_{\uparrow}^m z_{\downarrow}^j,
\end{aligned}
\eea
where \( \chi \) is obtained from Eq.~\ref{eq:chi_m} using \( M \), and \( M = \partial \ln(\mathcal{Z}) /  \partial (\beta h)  \), with $h = (\mu_\uparrow - \mu_\downarrow)/2$.

Using results up to fifth order, it is possible and beneficial to sum the virial expansion using the Pad\'e resummation method, as explained in Ref.~\cite{Hou2020}. Hence, for each physical quantity, we calculate the coefficients of a rational function of $z$,
\beq
F(z) = \frac{P_M(z)}{Q_K(z)} = \frac{p_0 + p_1 z + \cdots + p_M z^M}{1 + q_1 z + \cdots + q_K z^K},
\eeq
for each fixed value of $\beta h$ with $M+K = n_\text{max}$. Here, $n_\text{max} = 5$ is the maximum available order in the virial expansion. The choice of \( M \) and \( K \) is otherwise arbitrary but we follow the common practice of selecting \( M = K \) if possible or \( M = K + 1 \), i.e. we take either the diagonal or one-off-diagonal Pad\'e approximant. Note that the resummation~$F(z)$ is defined such that the virial expansion is reproduced by expanding~$F(z)$ up to order $n_\text{max}$ in $z$.

\section{Numerical results for the spin susceptibility}
In this section we present our results for the spin susceptibility as defined in~\equref{chi_m}. To put this quantity in dimensionless form, we use the Pauli susceptibility $\chi_{\mathrm{P}}^{} = 3 n / (2 \ef)$, which reflects the spin susceptibility of a noninteracting Fermi gas at $T=0$ where $\ef = \tf = \kf^2/2$, $\kf = (3\pi^2 n)^{1/3}$ and $n$ is the total density of the interacting system at a given value of $\mu = (\mu_\up + \mu_\down)/2$ and $h = (\mu_\up - \mu_\down)/2$. The spin-balanced limit corresponds to $\beta h = 0$, whereas values of $\beta h > 0$ reflect spin-polarized systems.

\begin{figure}[t]
  \centering
  \includegraphics{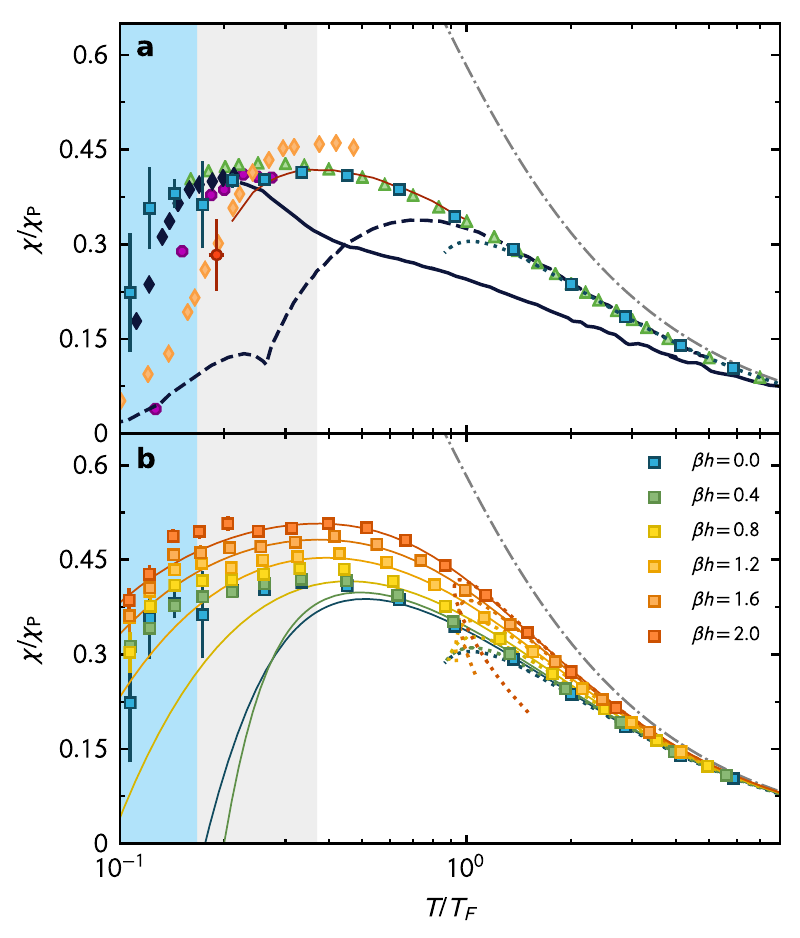}
  \caption{
  \label{fig:ufg_chi_ttf}
  Spin susceptibility $\chi$ in units of the Pauli susceptibility $\chi_{\mathrm{P}}^{}$ as a function of $T/\tf$. The blue (dark) shaded area shows the superfluid transition temperature in the balanced gas for orientation (with $T_c$ from experiment~\cite{Ku2012}), the gray (light) shaded area marks the maximum of the balanced curve (blue squares) and the dash-dotted line represents the noninteracting normalized susceptibility.
  (a) CL values (squares) for the balanced UFG are compared to and experimental value~\cite{Sanner2011} (red circle) as well as theoretical results from LW~\cite{Enss2012b} (green triangles), T-matrix~\cite{Palestini2012} (dark dashed line), extended T-matrix~\cite{Tajima2014} (thin red solid line), NSR~\cite{Pantel2014} (dark solid line), spherical cutoff AFQMC~\cite{Wlazlowski2013} (yellow diamonds) and two AFQMC studies~\cite{Jensen2019,Jensen2020} (dark diamonds) and ~\cite{Richie-Halford2020} (purple octagons).
  (b) CL results for the susceptibility for nonzero Zeeman field $\beta h = 0$ to $2.0$ (bottom to top lines, respectively) along with the corresponding values from the PRVE (thin solid lines). For comparison, the bare third-order VE is shown (dotted lines).
  }
\end{figure}

In panel (a) of~\figref{ufg_chi_ttf}, we show our results for the spin susceptibility as a function of $T/\tf$ for~$\beta h=0$ (blue squares). We find that, across all studied temperatures, the spin susceptibility is lower than that of the ideal Fermi gas (gray dashed-dotted line). At high temperature, we find excellent agreement with the third-order virial expansion (dotted line). At even higher temperature, our results tend to the expected $\sim 1/T$ decay according to Curie's law.

In addition to our CL results, a variety of other determinations of the spin susceptibility of the balanced UFG is shown in~\figref{ufg_chi_ttf}. Apart from the results from the Nozi\`eres-Schmitt-Rink (NSR) formalism~\cite{Pantel2014}, all theoretical approaches are found to be in agreement in the regime above $T/\tf \gtrsim 1$. In the regime \mbox{$T_c/\tf < T/\tf \lesssim 1$},
curves obtained from the NSR formalism~\cite{Pantel2014} as well as from T-matrix approaches~\cite{Palestini2012,Tajima2014} predict a substantial suppression of $\chi$.
Interestingly, also an AFQMC study in the grand-canonical ensemble~\cite{Wlazlowski2013} (yellow diamonds) predicts a substantial suppression of~$\chi$ below $T/\tf \approx 0.25$, which differs from a more recent AFQMC study at fixed particle number~\cite{Jensen2020}. This discrepancy was traced back to the lattice momentum cutoff which appears to leave a significant imprint in the results: while the earlier study relied on a spherical cutoff in momentum space, the latter considered the full Brillouin zone~\cite{Jensen2019,Jensen2020}. This was confirmed by the recent AFQMC study of Ref.~\cite{Richie-Halford2020}. Our CL results agree very well with the latter AFQMC studies as well as with results from studies based on the LW formalism~\cite{Enss2012b,Enss2012} across the entire temperature range, see~\figref{ufg_chi_ttf} for details.

The only experimental value~\cite{Sanner2011} shown in~\figref{ufg_chi_ttf} (depicted by the red circle in panel a) differs by roughly two standard deviations from the AFQMC results. Very recently, new experimental data for the spin susceptibility of the UFG became available~\cite{Long2020}. In the corresponding study, consistency with a mean-field approach was found, suggesting Fermi-liquid-type behavior. However, the presented values are trap-averaged such that a straightforward comparison to bulk-properties as presented here is not possible (see \appref{exp_compare} for a comparison of the integrated susceptibility). Finally, all theoretical values are at odds with a previous experimental determination of the spin susceptibility~\cite{Sommer2011} (not shown in the plot). This discrepancy is likely due to trap averaging in the analysis of the experimental data (see also Ref.~\cite{Enss2012}).

Among the aforementioned studies, the two recent AFQMC determinations, our CL values, and the LW results all display a very mild variation (on the order of a few percent at most) of $\chi/\chi_{\mathrm{P}}^{}$ as a function of $T/\tf$ in a region above $T_c$, before decreasing monotonically at high enough temperature. In other words, $\chi$ displays a very broad maximum centered at the spin-gap temperature $T_s > T_c$. Notably, these results clearly deviate from those for the noninteracting gas, which is a signal of strong interaction effects, even above the superfluid transition. Specifically, in the region around $T_s$ we have $\chi/\chi_{\mathrm{P}}^{}\gtrsim 0.6$ for the noninteracting system (for all the polarizations we considered; see below).

In panel (b) of~\figref{ufg_chi_ttf}, we show the spin susceptibility for polarized systems with $|\beta h| \le 2.0$. At very high temperature $T > \tf$, the curves for the polarized systems approach those from the virial expansion (dotted lines, color-coding as for CL) and eventually agree with the noninteracting limit, as in the balanced case. Moreover, at temperatures $T > T_c$, we find excellent agreement of our CL results at large $\beta h$ with those from the PRVE (thin solid lines, color-coding as for CL) up to fifth order, which is indeed not unexpected as interaction effects are more suppressed in that region. Said agreement deteriorates as the unpolarized limit is approached, where correlations are strongest. For $\beta h > 0$, we find that the functional form of $\chi/\chi_{\mathrm{P}}^{}$ as a function of $T/\tf$ is qualitatively the same as for the balanced system.
With increasing asymmetry $\beta h$, we observe a mild shift of the spin susceptibility toward larger values. As mentioned above, this is not unexpected as the system tends to evolve slowly towards the noninteracting gas with increasing polarization. However, even for $\beta h = 2.0$, the spin susceptibility of the interacting system still clearly differs from the one in the noninteracting limit.

Within the accuracy of our study, the aforementioned broad maximum in $\chi$ as a function of the temperature evolves to a plateau-like feature with increasing imbalance $\beta h$. Below the plateau we find that the spin susceptibility $\chi$ decreases rapidly with temperature, indicating the transition (or rather a crossover due to the finite system size) into the superfluid phase. Of course, the suppressed susceptibility by itself does not yet allow to make a statement on the phase transition to the superfluid phase since even a rapid suppression may be caused by, e.g., the formation of a pseudogap.
However, a peak in the compressibility of the UFG occurs at the temperatures associated with the lower end of the plateaus (for~$\beta h \gtrsim 1.2$)~\cite{Rammelmueller2018} which can be directly related to the transition temperature. Finally, we observe that, compared to the balanced case, the rapid decrease of the susceptibility sets in ``delayed'' when the temperature is lowered for increasing~$\beta h$. Together with the behavior of the compressibility, this indicates a decrease of the critical temperature with increasing imbalance (see also our discussion in Sec.~\ref{sec:pd} below).

We note that the occurrence of a plateau in $\chi$ above $T_c$ is not observed in studies of the attractive 2D Hubbard model where an ``anomalous suppression'' has been found~\cite{Trivedi1995,Randeria1992}. However, it should be pointed out that in such 2D models the effective interaction is much stronger than in the UFG, in the sense that in 2D a deep two-body bound state forms even in the absence of a background density. Similarly, Ref.~\cite{Richie-Halford2020} studied the 3D case for couplings stronger than the UFG (where a bound state is present at the two-body level) and found a clear increase in pairing correlations above the critical temperature~$T_c$.

From our analysis, we conclude that the UFG does not show an anomalously suppressed spin susceptibility for imbalances $\beta h \gtrsim 1.2$ in the sense of the attractive 2D Hubbard model. For $\beta h \lesssim 1.2$, the UFG may feature a pseudogap region but its precise limits and characterization can not be conclusively studied with thermodynamic probes (e.g., the spin susceptibility) used in our present work. In the following, we will relate these findings, as far as possible, to the phase diagram of the spin-polarized UFG.

\section{Implications for the phase diagram}\label{sec:pd}
Despite intense theoretical and experimental investigations during the past two decades, our understanding of the full phase diagram of the polarized UFG is still incomplete. With respect to the overall structure of the phase diagram, we summarize main results obtained from non-perturbative studies beyond the mean-field approximation~\cite{Parish2007}. A more complete discussion of the precise nature of the superfluid phase, including potentially occurring exotic superfluids, is beyond this work (see, e.g., Refs.~\cite{Bulgac2006,Radzihovsky2010,Chevy2010,Gubbels2013} for reviews).

\subsection{Overview on the phase structure}
\begin{figure*}
  \includegraphics{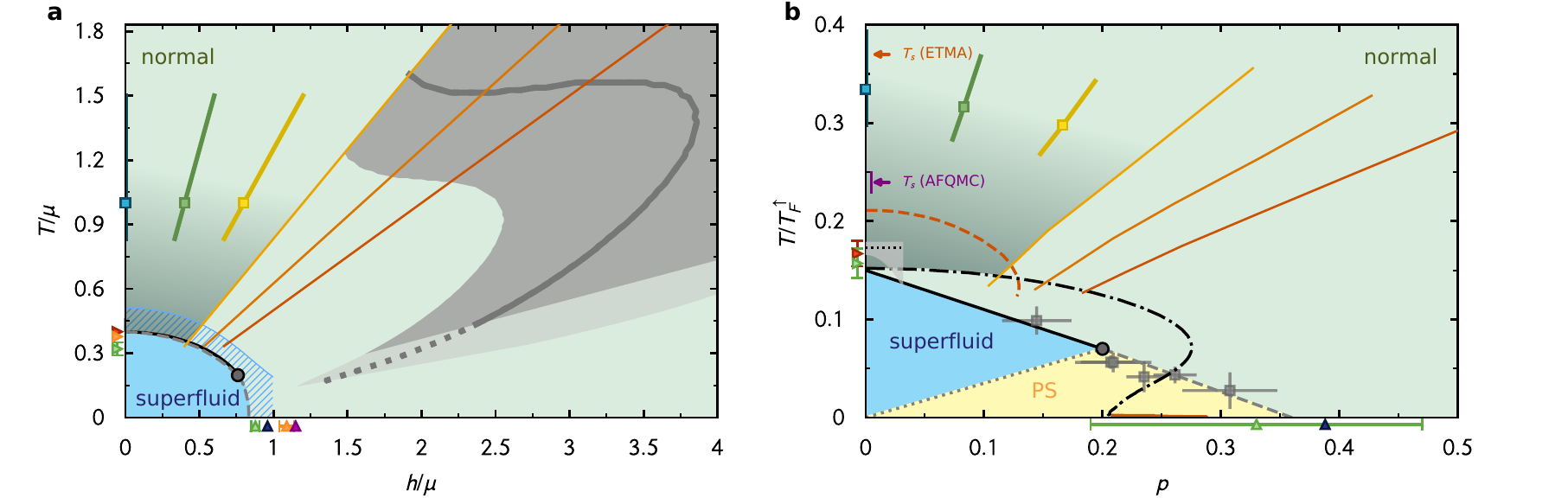}
  \caption{\label{fig:ufg_phase_diagrams} Phase diagrams of the spin-polarized UFG.
  (a) Phase diagram spanned by the dimensionless temperature $(\beta\mu)^{-1}$ and the dimensionless Zeeman field $h/\mu$. The black solid line depicts the second-order phase transition line from a fRG study~\cite{Boettcher2014} where the black dot represents the location of the critical point. The thick line reflects the spin-gap temperature $T_s$ as obtained from the PRVE along with an uncertainty estimate (shaded area, see main text).
  (b) Phase diagram spanned by~$T/\tf[\up]$ and the polarization as measured in experiment~\cite{Shin2008} (gray squares reflect experimental measurements, thick lines are the inferred phase boundaries) and compared to recent determinations via LW theory~\cite{Frank2018,Pini2021} (dot-dashed line) as well as ETMA~\cite{Tajima2014} (red dashed line). For the balanced limit $T_s$ is shown as determined via AFQMC~\cite{Richie-Halford2020} and ETMA~\cite{Tajima2014}.
  In both panels, squares reflect the maximum of the CL data and thick lines indicate the uncertainty estimate (see main text). The thin colored lines (color coding as in \figref{ufg_chi_ttf}) reflect the the observed plateau in the spin-susceptibility ($\beta h = 1.2, 1.6, 2.0$ from left to right, respectively). Critical values correspond to experimental values from the MIT group~\cite{Ku2012}~(red triangles) and ENS group~\cite{Nascimbene2010,Nascimbene2011}~(green triangles), LW results~\cite{Frank2018}~(orange triangle), FN-DMC calculations~\cite{Lobo2006}~(dark blue triangles), $\epsilon$-expansion (purple triangle) and Worm-MC data~\cite{Goulko2010} (light gray-shaded area) [see \appref{lit_data} for the corresponding numerical data]. The black shaded areas in both panels mark potential pseudogap regions (see main text).}
\end{figure*}

In \figref{ufg_phase_diagrams} we consider the phase diagram in two different forms: Panel (a) depicts the phase diagram as spanned by the temperature $T$ measured in units of the chemical potential~$\mu$, and the so-called Zeeman field~$h$ also given in units of~$\mu$. The color coding reflects the phase diagram as obtained from a functional renormalization group (fRG) study~\cite{Boettcher2014}, the black solid line depicts the corresponding second-order phase transition line between the normal and the superfluid phase which terminates in a critical point (black dot) where it meets a first-order transition line (see also Refs.~\cite{Birse2005,Diehl2007} for early studies of non-relativistic Fermi gases with this approach and Refs.~\cite{Diehl2009, Scherer2010, Boettcher2012} for corresponding reviews).
In the balanced limit ($h/\mu = 0$), the fRG result for the critical temperature result agrees well with two experimental determinations of $(T/\mu)_c\approx0.4$~\cite{Nascimbene2010,Ku2012}. We add that more recent studies based on LW formalism also find this value for $(T/\mu)_c$ in the balanced limit~\cite{Frank2018,Pini2019}.

While $T_c$ in the balanced case is well understood~\cite{MCNote}, much less is known about the critical field strength~$(h/\mu)_c$ at which the system undergoes a transition from the superfluid phase to the normal phase in the zero-temperature limit. This transition is expected to be of first order~\cite{Radzihovsky2010,Chevy2010,Gubbels2013,Boettcher2014,Frank2018}, see also Ref.~\cite{Zdybel2020} for a recent discussion of quantum Lifshitz points and fluctuation-induced first-order phase transitions in imbalanced Fermi mixtures.
In this regime of the phase diagram, the results from fRG~\cite{Boettcher2014}, experiment~\cite{Nascimbene2010} and FN-DMC~\cite{Lobo2006} are in accordance (although not in full quantitative agreement) but disagree with those from LW theory~\cite{Frank2018} and the $\epsilon$-expansion~\cite{Nishida2007b} which find a significantly higher critical value $(h/\mu)_c$.

In panel (b) of~\figref{ufg_phase_diagrams}, an experimental measurement of the phase diagram (color coding as in panel a) is shown in the plane spanned by the temperature~$T$ in units of the Fermi temperature~$\tf[\up]$ and the polarization~$p=(N_{\uparrow}-N_{\downarrow})/N_{\uparrow}+N_{\downarrow})$. The solid black line represents an estimate for the phase transition line as extracted from an analysis of experimental data for a trapped gas based a local density approximation, see Ref.~\cite{Shin2008} for details. The black dash-dotted line reflects the critical line from LW theory~\cite{Frank2018} and a more recent fully self-consistent T-matrix calculation~\cite{Pini2021} (both approaches are, in fact, identical) which agree well with experimental data at $|p|=0$. Although different from the experimental determination at $p>0$, we add that the flatness of the phase transition line at small polarization observed in the LW studies is in accordance with Monte Carlo studies based on the Worm algorithm~\cite{Goulko2010} which cover polarizations $|p|\lesssim 0.035$ (gray shaded area). The latter study is the only stochastic determination of $T_c$ beyond $p = 0$ to date, to the best of our knowledge. Such a weak dependence of the phase transition line on the polarization is also suggested by a recent CL study of the compressibility~\cite{Rammelmueller2018}. Note that there are further studies of the phase structure of the UFG based on a range of methods. For example, determinations of the second-order transition line based on NSR~\cite{Pantel2014} and extended T-matrix approximation~\cite{Kashimura2012} (ETMA, red dashed line). Both predict a $T_c$ for the balanced gas well above $T/\tf[\up] = 0.2$. Moreover, an early RG study reported quantitative agreement with the experimental value for the critical temperature in the balanced case as well as for the location of the critical point~\cite{Gubbels2013}, see also Ref.~\cite{Strinati2018} for a discussion.

At zero temperature, the experimental determination of the critical polarization~$p_c$ by the ENS group exhibits a large uncertainty~\cite{Nascimbene2010} and does therefore not allow to discriminate between state-of-the-art predictions for this quantity from theory. However, the latter are all well below the mean-field estimate which is~$p_c \approx 0.93$ \cite{Gubbels2013}.
\subsection{The pseudogap regime}
In addition to the occurrence of (potentially inhomogeneous) superfluid phases at low temperatures, pre-formed pairs in the normal phase could lead to non-Fermi liquid behavior in some regions of the phase diagram, which we refer to as ``pseudogap regime'' in the following. Here, we relate our results for the spin susceptibility~$\chi$ to various regions in the phase diagram by extracting an estimate for the spin-gap temperature $T_s$. The latter is defined to be the maximum of $\chi/\chi_{\mathrm{P}}^{}$, see \figref{ufg_chi_ttf} for our results for $\chi/\chi_{\mathrm{P}}^{}$ as a function of the temperature for various values of the Zeeman field~$h$. As mentioned above, $\chi/\chi_{\mathrm{P}}^{}$ exhibits a relatively broad maximum, which complicates a precise determination of its position. Nevertheless, a search for such a maximum allows us to identify the region where $\chi/\chi_{\mathrm{P}}^{}$ starts to display an anomalous suppression, potentially related to the onset of pairing effects.

We begin by analyzing our CL data for $|\beta h|\lesssim 0.8$. The maximum of the spin-susceptibility obtained with CL is shown as filled squares in both panels of \figref{ufg_phase_diagrams}. Since the resolution of possible peak-position is limited by the grid of input parameters to the simulation (we chose a constant spacing of $\Delta\beta\mu = 0.5$ for fixed $\beta h$), the true maximum could lie between those points. In order to give an estimate of the uncertainty of the true maximum we have indicated half the distance to the neighboring data points with thick lines.
Note that this estimate does not account for statistical and systematic errors in our CL study. In panel (b) of the figure, the original determination of $T_s$ for the balanced gas from an ETMA calculation~\cite{Tajima2014} as well as from a recent AFQMC study~\cite{Richie-Halford2020} are shown for comparison. While the spin-gap temperature from the former study is found to be higher than the one presented here (although within the above described error), it is found to be lower in the AFQMC calculations.

As mentioned above, for $1.2\lesssim \beta h\lesssim 2.0$, the CL data for the normalized spin susceptibility does not show a distinct maximum but is rather characterized by a broad plateau. Towards lower temperatures, we then observe a rapid decrease which indicates the onset of the superfluid phase transition. Given the accuracy of our present data, we refrain from indicating the positions of the maxima of $\chi/\chi_{\mathrm{P}}^{}$ by filled squares as we do for $|\beta h|\lesssim 0.8$ but only represent the plateau as solid lines in both panels of \figref{ufg_phase_diagrams}. The lower ends of these lines correspond to the last data point before the aforementioned rapid decrease at lower temperatures sets in. In the phase diagram spanned by the temperature and polarization (panel (b) of \figref{ufg_phase_diagrams}), we observe that the lower ends of the plateaus for $\beta h = 1.2, 1.6$, and $2.0$ overlap with the superfluid phase boundary obtained from LW theory~\cite{Frank2018}. Taking at face value the position of the phase boundary obtained from LW theory, our CL results suggest that there is no room for a regime with an anomalously suppressed spin susceptibility in this region of the phase diagram. We emphasize that the lines associated with the plateaus for $\beta h = 1.2, 1.6$, and $2.0$ should not be confused with the thick solid lines for $|\beta h|\lesssim 0.8$ which indicate an estimate for the uncertainty of the spin-gap temperature~$T_s$.

Similarly, in the phase diagram spanned by the temperature and the Zeeman field (panel (a) of \figref{ufg_phase_diagrams}), the lines associated with the plateaus of the spin susceptibility for $1.2\leq \beta h\leq 2.0$ extend down to the phase boundary obtained from a fRG study~\cite{Boettcher2014}. In the latter study a precondensation regime has been identified (blue hatched band). In this regime, the gap vanishes in the long-range limit but ``local ordering'' is still present on intermediate length scales, leading to a strong suppression of the density of states above the superfluid transition. As discussed in Ref.~\cite{Boettcher2014}, this precondensation regime may be used to estimate the emergence of a pseudogap region in the phase diagram. However, the definition of this regime cannot be straightforwardly related to the definition of the pseudogap regime based on the spin susceptibility underlying our present analysis.

In addition to the CL results above, we have performed the same analysis for the PRVE. The position of the maximum of $\chi/\chi_{\mathrm{P}}^{}$ as a function of $T/\tf$ is indicated by the thick line (continuous and dashed, see below) in \figref{ufg_phase_diagrams} (panel a). The corresponding uncertainty estimate (dark and light shaded area, see below) reflects the distance between the points where $\chi/\chi_{\mathrm{P}}^{}$ is $99.7\%$ of the maximum value. The solid thick line and corresponding dark shaded area reflect the parameter region where we expect the PRVE to be quantitatively trustworthy; in particular, part of the dark shaded area overlaps (and agrees with) the CL results described above. The dashed thick line and corresponding light shaded area display the parameter region where resummation techniques begin to fail due to strong interaction effects, in particular at low polarizations.

Judging from our analysis of the spin susceptibility in the light of other phase-diagram studies, we presently expect that a notable pseudogap regime may only develop in the regime $\beta h \lesssim 1.2$, if present at all. In particular, we do not observe a pseudogap regime of the form reported for the attractive Hubbard model~\cite{Trivedi1995,Randeria1992}). Our data is compatible with the picture of a narrow ``pseudogap band'' above the superfluid phase which continuously shrinks as $\beta h$ is increased until it eventually disappears. In this scenario one could interpret the observation of the plateau-like feature in the susceptibility for $\beta h \gtrsim 1.2$ as the absence of a pseudogap regime in this part of the phase diagram. The black shaded areas in both panels of \figref{ufg_phase_diagrams} highlight the most-likely regions in which a pseudogap may develop, based on our findings.

\section{Discussion \& outlook}
To summarize, we have presented a numerical calculation of the magnetic susceptibility for the spin-imbalanced UFG based on the CL approach. In the balanced case, our results agree with several previous results from the literature across the entire temperature range in the normal phase. For the spin-polarized case, we validate the accuracy of these CL results with a PRVE up to fifth order, with good agreement ranging from high temperatures until well below $\tf$. Furthermore, we use this quantity as a probe for the influence of pairing in the normal phase. Our data shows a slight suppression of $\chi/\chi_{\mathrm{P}}^{}$ for small imbalances with a spin-gap temperature well above $T_c$. A strong suppression, as it is for instance observed in 2D Hubbard models superconductors~\cite{Trivedi1995,Randeria1992}, is not observed. Nevertheless, the occurrence of a pseudogap regime for smaller imbalances ($\beta h \lesssim 1.2$) cannot be ruled out with our present study.

For $\beta h \gtrsim 1.2$, we report a plateau-like feature of the spin susceptibility just above the phase transition and a subsequent abrupt drop at low temperatures. Such a suppression is expected to eventually occur in the low-temperature limit due to the formation of a pairing gap.  Notably, we find that the onset of the rapid decrease of the normalized spin susceptibility $\chi/\chi_{\mathrm{P}}^{}$ is shifted towards lower temperatures when~$\beta h$ is increased which suggests a decrease of the critical temperature for superfluidity with increased imbalance, in agreement with previous studies of the phase diagram. In our case, this interpretation is supported by the fact that the position of the lower end of the plateau in the susceptibility coincides with the peak of the compressibility~\cite{Rammelmueller2018}. However, for a rigorous determination of the $\beta h$-dependence of the critical temperature, a finite-size scaling analysis as well as an extrapolation to the zero-density limit (as, e.g., in Ref.~\cite{Jensen2019b}) should be performed.

The shrinking of a potential pseudogap regime with increasing~$\beta h$ deduced from our analysis of the spin susceptibility supports the picture that the nature of the normal phase above the critical polarization at low temperatures is well described by Fermi liquid theory. This has been previously conjectured based on qualitative results from a T-matrix study~\cite{Mueller2011} as well as FN-DMC~\cite{Lobo2006} and also agrees with experimental findings for large imbalances at low temperatures~\cite{Nascimbene2010,Nascimbene2010b,Nascimbene2011}. It must be noted, however, that our present study cannot be conclusive in this regard as our analysis solely relies on thermodynamic observables.

In summary, our findings further corroborate that thermodynamic quantities are not ideal probes for pseudogap physics in the UFG. In particular, these results show that an analysis of the spin susceptibility and the associated spin-gap temperature is not sufficient by itself to fully understand pairing effects, low energy excitations and the potential formation of a pseudogap regime above the critical temperature.

As a consequence for experiments, this suggests that a more complete characterization of a potential pseudogap regime is likely best achieved by probing the spectral function as in Ref.~\cite{Stewart2008,Gaebler2010,Sagi2015}, where a back-bending of the dispersion relation near $\kf$ would indicate pairing effects~\cite{Zwerger2012}. However, these experiments are subject to ambiguous interpretation and consensus has yet to be reached.

Perhaps a combination with the recent development of flat trap geometries in ultracold Fermi gases ~\cite{Mukherjee2016,Hueck2017} will lead to a more complete understanding of the UFG above the superfluid transition. Moreover, modern spectroscopic techniques, which allowed to discern between two- and many-body pairing in 2D Fermi gases~\cite{Murthy2018}, could shed some more light on this matter. Finally, we note that a viable alternative to spectroscopic probes could be the investigation of density-density correlations in momentum space, the so-called shot-noise, which exhibits distinct signatures of pairing~\cite{Altman2004,Rammelmueller2020}. This is work in progress.


\acknowledgments
The authors would like to thank Tilman Enss, Yoram Alhassid, Scott Jensen and Michele Pini for useful comments on the manuscript.
J.B. acknowledges support by the DFG under grant BR 4005/4-1 (Heisenberg program) and BR 4005/5-1. J.B. and L.R. acknowledge support by HIC for FAIR within the LOEWE program of the State of Hesse. This
material is based upon work supported by the National Science Foundation under Grants No. PHY{1452635} and PHY{2013078}. Numerical calculations have been performed on the LOEWE-CSC Frankfurt.\\

\appendix

\section{\label{app:exp_compare} Comparison of integrated susceptibility}
\begin{figure}
  \includegraphics{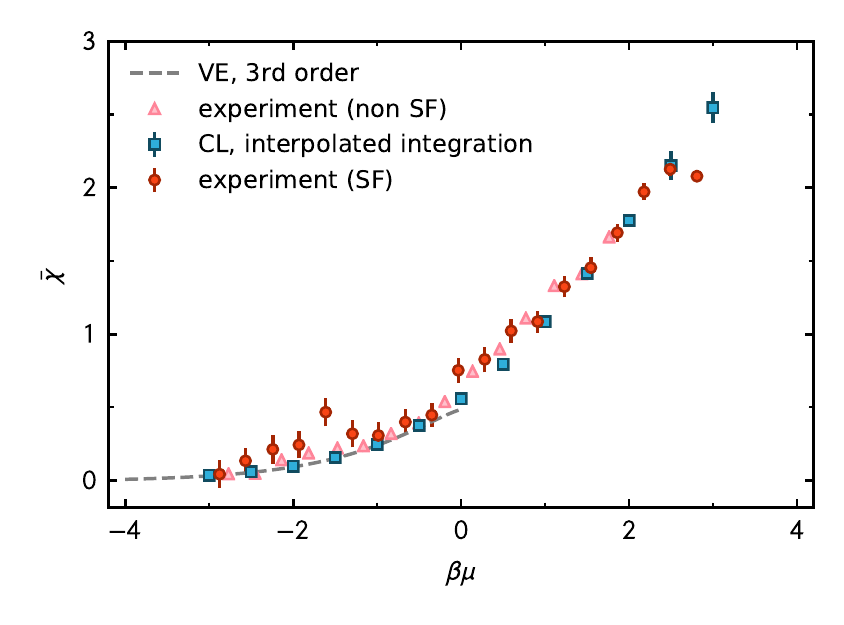}
  \caption{\label{fig:ufg_exp_comparison} Integrated susceptibility $\bar{\chi}$ as numerically obtained in this work (blue squares) compared to a recent experimental measurement (red circles and pink triangles). For low fugacity (large negative $\beta\mu$), also the third-order virial expansion is shown (gray dashed line).}
\end{figure}

As mentioned in the main text, a recent experimental study~\cite{Long2020} presented trap-averaged data for the spin-susceptibility of the spin-balanced UFG. For the bulk system, as studied here via periodic boundary conditions, also the integrated spin-susceptibility
\begin{equation}
 \bar{\chi} = \int_{-\infty}^{\mu}\d\mu'\ \chi(\mu')\lambda_T^3
\end{equation}
was measured. We show a comparison of our data in the normal phase (i.e., $\beta\mu \lesssim 2.5$) to this measurement in \figref{ufg_exp_comparison}. We observe overall good agreement. The errorbars of the CL values reflect linearly propagated statistical errors from the bare data.

\section{Literature data for critical values~\label{app:lit_data}}
In~\tabref{ufg_numbers}, we summarize values for the critical polarization $p_c$ and the critical magnetic field $h_c$ (in units of $\mu$) at zero temperature together with results for the critical temperature $T_c$, the spin-gap temperature $T_s$ and the related pairing temperature $T^*$ in the balanced limit.

\begin{table*}
  \centering
  \begin{small}
    \begin{tabular}{@{}L{0.18\textwidth}L{0.2\textwidth}M{0.06\textwidth}M{0.08\textwidth}M{0.08\textwidth}M{0.08\textwidth}M{0.08\textwidth}M{0.08\textwidth}M{0.08\textwidth}@{}N}\toprule
      {\bf technique} & {\bf authors} & {\bf year} & $T_c/\tf$ & $p_c$ & $(\beta\mu)_c$ &  $(\frac{h}{\mu})_c$ & $T^*/\tf$ &$T_s/\tf$ & \\[8pt]\toprule
      experiment (MIT) & Shin \etal ~\cite{Shin2008} & 2008 & $\sim 0.15$ & $\sim 0.36$ & --- & $\sim 0.95$ & --- & --- & \\[8pt]
      experiment (ENS) & Nascimbene \etal ~\cite{Nascimbene2010,Nascimbene2011} & 2010 & $0.157(15)$ & $0.33(14)$ & $3.13(29)$ & $0.878(28)$& --- & --- & \\[8pt]
      experiment (MIT) & Ku \etal ~\cite{Ku2012} & 2012 & $0.167(13)$ & --- & $\sim 2.5 $& --- & --- & --- & \\[8pt] \midrule
      FN-DMC & Lobo \etal~\cite{Lobo2006} & 2006 & --- & $0.388$ & --- & $0.96$ & --- & --- & \\[8pt]
      $\epsilon$-expansion & Nishida \etal ~\cite{Nishida2007b} & 2007 & --- & --- & --- & $1.15$ & --- & --- & \\[8pt]
      extended T-matrix & Tajima \etal~\cite{Tajima2014} & 2014 & 0.21 & --- & --- &  ---  & --- & 0.37 & \\[8pt]
      functional RG (fRG) & Boettcher \etal~\cite{Boettcher2014} & 2015 & --- & --- & $2.5$ & $0.83$ & --- & --- & \\[8pt]
      Luttinger-Ward & Frank \etal~\cite{Frank2018} & 2018 &  $0.152$ & --- & $2.65$ & $1.09(5)$ & --- & --- & \\[8pt]
      fully self-consistent T-matrix & Pini \etal~\cite{Pini2019} & 2019 &  $0.1505$ & --- & $2.658$ & --- & 0.183(2) & --- & \\[8pt]
      AFQMC & Richie-Halford \etal~\cite{Richie-Halford2020} & 2020 & $0.16(2)$ & --- & --- &  ---  & $0.21(3)$ & $ >0.24(1)$ & \\[8pt]
      CL & this work & 2021 & --- & --- & --- &  ---  & --- & $0.33(5)$ & \\[8pt] \toprule
    \end{tabular}    
  \end{small}
  \caption{\label{tab:ufg_numbers} Selected list of values of critical quantities of the UFG   in the grand-canonical and canonical ensembles. Uncertainties are indicated, where available. Note that the results for the critical temperature $T_c$, the spin-gap temperature $T_s$ and the related pairing temperature $T^*$ are the ones for the balanced UFG. Acronyms: fixed-node diffusion Monte Carlo (FN-DMC) and auxiliary-field Monte Carlo (AFQMC).}
\end{table*}

\bibliography{pseudogap_refs}

\begin{thebibliography}{101}%
\makeatletter
\providecommand \@ifxundefined [1]{%
 \@ifx{#1\undefined}
}%
\providecommand \@ifnum [1]{%
 \ifnum #1\expandafter \@firstoftwo
 \else \expandafter \@secondoftwo
 \fi
}%
\providecommand \@ifx [1]{%
 \ifx #1\expandafter \@firstoftwo
 \else \expandafter \@secondoftwo
 \fi
}%
\providecommand \natexlab [1]{#1}%
\providecommand \enquote  [1]{``#1''}%
\providecommand \bibnamefont  [1]{#1}%
\providecommand \bibfnamefont [1]{#1}%
\providecommand \citenamefont [1]{#1}%
\providecommand \href@noop [0]{\@secondoftwo}%
\providecommand \href [0]{\begingroup \@sanitize@url \@href}%
\providecommand \@href[1]{\@@startlink{#1}\@@href}%
\providecommand \@@href[1]{\endgroup#1\@@endlink}%
\providecommand \@sanitize@url [0]{\catcode `\\12\catcode `\$12\catcode
  `\&12\catcode `\#12\catcode `\^12\catcode `\_12\catcode `\%12\relax}%
\providecommand \@@startlink[1]{}%
\providecommand \@@endlink[0]{}%
\providecommand \url  [0]{\begingroup\@sanitize@url \@url }%
\providecommand \@url [1]{\endgroup\@href {#1}{\urlprefix }}%
\providecommand \urlprefix  [0]{URL }%
\providecommand \Eprint [0]{\href }%
\providecommand \doibase [0]{https://doi.org/}%
\providecommand \selectlanguage [0]{\@gobble}%
\providecommand \bibinfo  [0]{\@secondoftwo}%
\providecommand \bibfield  [0]{\@secondoftwo}%
\providecommand \translation [1]{[#1]}%
\providecommand \BibitemOpen [0]{}%
\providecommand \bibitemStop [0]{}%
\providecommand \bibitemNoStop [0]{.\EOS\space}%
\providecommand \EOS [0]{\spacefactor3000\relax}%
\providecommand \BibitemShut  [1]{\csname bibitem#1\endcsname}%
\let\auto@bib@innerbib\@empty
\bibitem [{\citenamefont {Perali}\ \emph {et~al.}(2002)\citenamefont {Perali},
  \citenamefont {Pieri}, \citenamefont {Strinati},\ and\ \citenamefont
  {Castellani}}]{Perali2002}%
  \BibitemOpen
  \bibfield  {author} {\bibinfo {author} {\bibfnamefont {A.}~\bibnamefont
  {Perali}}, \bibinfo {author} {\bibfnamefont {P.}~\bibnamefont {Pieri}},
  \bibinfo {author} {\bibfnamefont {G.~C.}\ \bibnamefont {Strinati}},\ and\
  \bibinfo {author} {\bibfnamefont {C.}~\bibnamefont {Castellani}},\ }\bibfield
   {title} {\bibinfo {title} {Pseudogap and spectral function from
  superconducting fluctuations to the bosonic limit},\ }\href
  {https://doi.org/10.1103/PhysRevB.66.024510} {\bibfield  {journal} {\bibinfo
  {journal} {Phys. Rev. B}\ }\textbf {\bibinfo {volume} {66}},\ \bibinfo
  {pages} {024510} (\bibinfo {year} {2002})}\BibitemShut {NoStop}%
\bibitem [{\citenamefont {Chen}\ \emph {et~al.}(2005)\citenamefont {Chen},
  \citenamefont {Stajic}, \citenamefont {Tan},\ and\ \citenamefont
  {Levin}}]{Chen2005}%
  \BibitemOpen
  \bibfield  {author} {\bibinfo {author} {\bibfnamefont {Q.}~\bibnamefont
  {Chen}}, \bibinfo {author} {\bibfnamefont {J.}~\bibnamefont {Stajic}},
  \bibinfo {author} {\bibfnamefont {S.}~\bibnamefont {Tan}},\ and\ \bibinfo
  {author} {\bibfnamefont {K.}~\bibnamefont {Levin}},\ }\bibfield  {title}
  {\bibinfo {title} {Bcs-bec crossover: From high temperature superconductors
  to ultracold superfluids},\ }\href
  {https://doi.org/https://doi.org/10.1016/j.physrep.2005.02.005} {\bibfield
  {journal} {\bibinfo  {journal} {Phys. Rep.}\ }\textbf {\bibinfo {volume}
  {412}},\ \bibinfo {pages} {1 } (\bibinfo {year} {2005})}\BibitemShut
  {NoStop}%
\bibitem [{\citenamefont {Bulgac}\ \emph {et~al.}(2006)\citenamefont {Bulgac},
  \citenamefont {Forbes},\ and\ \citenamefont {Schwenk}}]{Bulgac2006}%
  \BibitemOpen
  \bibfield  {author} {\bibinfo {author} {\bibfnamefont {A.}~\bibnamefont
  {Bulgac}}, \bibinfo {author} {\bibfnamefont {M.~M.}\ \bibnamefont {Forbes}},\
  and\ \bibinfo {author} {\bibfnamefont {A.}~\bibnamefont {Schwenk}},\
  }\bibfield  {title} {\bibinfo {title} {Induced $p$-wave superfluidity in
  asymmetric fermi gases},\ }\href
  {https://doi.org/10.1103/PhysRevLett.97.020402} {\bibfield  {journal}
  {\bibinfo  {journal} {Phys. Rev. Lett.}\ }\textbf {\bibinfo {volume} {97}},\
  \bibinfo {pages} {020402} (\bibinfo {year} {2006})}\BibitemShut {NoStop}%
\bibitem [{\citenamefont {Stewart}\ \emph {et~al.}(2008)\citenamefont
  {Stewart}, \citenamefont {Gaebler},\ and\ \citenamefont {Jin}}]{Stewart2008}%
  \BibitemOpen
  \bibfield  {author} {\bibinfo {author} {\bibfnamefont {J.~T.}\ \bibnamefont
  {Stewart}}, \bibinfo {author} {\bibfnamefont {J.~P.}\ \bibnamefont
  {Gaebler}},\ and\ \bibinfo {author} {\bibfnamefont {D.~S.}\ \bibnamefont
  {Jin}},\ }\bibfield  {title} {\bibinfo {title} {Using photoemission
  spectroscopy to probe a strongly interacting fermi gas},\ }\href
  {https://doi.org/10.1038/nature07172} {\bibfield  {journal} {\bibinfo
  {journal} {Nature}\ }\textbf {\bibinfo {volume} {454}},\ \bibinfo {pages}
  {744} (\bibinfo {year} {2008})}\BibitemShut {NoStop}%
\bibitem [{\citenamefont {Kuhnle}\ \emph {et~al.}(2010)\citenamefont {Kuhnle},
  \citenamefont {Hu}, \citenamefont {Liu}, \citenamefont {Dyke}, \citenamefont
  {Mark}, \citenamefont {Drummond}, \citenamefont {Hannaford},\ and\
  \citenamefont {Vale}}]{Kuhnle2010}%
  \BibitemOpen
  \bibfield  {author} {\bibinfo {author} {\bibfnamefont {E.~D.}\ \bibnamefont
  {Kuhnle}}, \bibinfo {author} {\bibfnamefont {H.}~\bibnamefont {Hu}}, \bibinfo
  {author} {\bibfnamefont {X.-J.}\ \bibnamefont {Liu}}, \bibinfo {author}
  {\bibfnamefont {P.}~\bibnamefont {Dyke}}, \bibinfo {author} {\bibfnamefont
  {M.}~\bibnamefont {Mark}}, \bibinfo {author} {\bibfnamefont {P.~D.}\
  \bibnamefont {Drummond}}, \bibinfo {author} {\bibfnamefont {P.}~\bibnamefont
  {Hannaford}},\ and\ \bibinfo {author} {\bibfnamefont {C.~J.}\ \bibnamefont
  {Vale}},\ }\bibfield  {title} {\bibinfo {title} {Universal behavior of pair
  correlations in a strongly interacting fermi gas},\ }\href
  {https://doi.org/10.1103/PhysRevLett.105.070402} {\bibfield  {journal}
  {\bibinfo  {journal} {Phys. Rev. Lett.}\ }\textbf {\bibinfo {volume} {105}},\
  \bibinfo {pages} {070402} (\bibinfo {year} {2010})}\BibitemShut {NoStop}%
\bibitem [{\citenamefont {Perali}\ \emph {et~al.}(2011)\citenamefont {Perali},
  \citenamefont {Palestini}, \citenamefont {Pieri}, \citenamefont {Strinati},
  \citenamefont {Stewart}, \citenamefont {Gaebler}, \citenamefont {Drake},\
  and\ \citenamefont {Jin}}]{Perali2011}%
  \BibitemOpen
  \bibfield  {author} {\bibinfo {author} {\bibfnamefont {A.}~\bibnamefont
  {Perali}}, \bibinfo {author} {\bibfnamefont {F.}~\bibnamefont {Palestini}},
  \bibinfo {author} {\bibfnamefont {P.}~\bibnamefont {Pieri}}, \bibinfo
  {author} {\bibfnamefont {G.~C.}\ \bibnamefont {Strinati}}, \bibinfo {author}
  {\bibfnamefont {J.~T.}\ \bibnamefont {Stewart}}, \bibinfo {author}
  {\bibfnamefont {J.~P.}\ \bibnamefont {Gaebler}}, \bibinfo {author}
  {\bibfnamefont {T.~E.}\ \bibnamefont {Drake}},\ and\ \bibinfo {author}
  {\bibfnamefont {D.~S.}\ \bibnamefont {Jin}},\ }\bibfield  {title} {\bibinfo
  {title} {Evolution of the normal state of a strongly interacting fermi gas
  from a pseudogap phase to a molecular bose gas},\ }\href
  {https://doi.org/10.1103/PhysRevLett.106.060402} {\bibfield  {journal}
  {\bibinfo  {journal} {Phys. Rev. Lett.}\ }\textbf {\bibinfo {volume} {106}},\
  \bibinfo {pages} {060402} (\bibinfo {year} {2011})}\BibitemShut {NoStop}%
\bibitem [{\citenamefont {Palestini}\ \emph {et~al.}(2012)\citenamefont
  {Palestini}, \citenamefont {Pieri},\ and\ \citenamefont
  {Strinati}}]{Palestini2012}%
  \BibitemOpen
  \bibfield  {author} {\bibinfo {author} {\bibfnamefont {F.}~\bibnamefont
  {Palestini}}, \bibinfo {author} {\bibfnamefont {P.}~\bibnamefont {Pieri}},\
  and\ \bibinfo {author} {\bibfnamefont {G.~C.}\ \bibnamefont {Strinati}},\
  }\bibfield  {title} {\bibinfo {title} {Density and spin response of a
  strongly interacting fermi gas in the attractive and quasirepulsive regime},\
  }\href {https://link.aps.org/doi/10.1103/PhysRevLett.108.080401} {\bibfield
  {journal} {\bibinfo  {journal} {Phys. Rev. Lett.}\ }\textbf {\bibinfo
  {volume} {108}},\ \bibinfo {pages} {080401} (\bibinfo {year}
  {2012})}\BibitemShut {NoStop}%
\bibitem [{\citenamefont {Enss}\ and\ \citenamefont
  {Haussmann}(2012)}]{Enss2012b}%
  \BibitemOpen
  \bibfield  {author} {\bibinfo {author} {\bibfnamefont {T.}~\bibnamefont
  {Enss}}\ and\ \bibinfo {author} {\bibfnamefont {R.}~\bibnamefont
  {Haussmann}},\ }\bibfield  {title} {\bibinfo {title} {Quantum mechanical
  limitations to spin diffusion in the unitary fermi gas},\ }\href
  {https://link.aps.org/doi/10.1103/PhysRevLett.109.195303} {\bibfield
  {journal} {\bibinfo  {journal} {Phys. Rev. Lett.}\ }\textbf {\bibinfo
  {volume} {109}},\ \bibinfo {pages} {195303} (\bibinfo {year}
  {2012})}\BibitemShut {NoStop}%
\bibitem [{\citenamefont {Boettcher}\ \emph {et~al.}(2014)\citenamefont
  {Boettcher}, \citenamefont {Pawlowski},\ and\ \citenamefont
  {Wetterich}}]{Boettcher2013}%
  \BibitemOpen
  \bibfield  {author} {\bibinfo {author} {\bibfnamefont {I.}~\bibnamefont
  {Boettcher}}, \bibinfo {author} {\bibfnamefont {J.~M.}\ \bibnamefont
  {Pawlowski}},\ and\ \bibinfo {author} {\bibfnamefont {C.}~\bibnamefont
  {Wetterich}},\ }\bibfield  {title} {\bibinfo {title} {Critical temperature
  and superfluid gap of the unitary fermi gas from functional
  renormalization},\ }\href {https://doi.org/10.1103/PhysRevA.89.053630}
  {\bibfield  {journal} {\bibinfo  {journal} {Phys. Rev. A}\ }\textbf {\bibinfo
  {volume} {89}},\ \bibinfo {pages} {053630} (\bibinfo {year}
  {2014})}\BibitemShut {NoStop}%
\bibitem [{\citenamefont {Wlaz\l{}owski}\ \emph {et~al.}(2013)\citenamefont
  {Wlaz\l{}owski}, \citenamefont {Magierski}, \citenamefont {Drut},
  \citenamefont {Bulgac},\ and\ \citenamefont {Roche}}]{Wlazlowski2013}%
  \BibitemOpen
  \bibfield  {author} {\bibinfo {author} {\bibfnamefont {G.}~\bibnamefont
  {Wlaz\l{}owski}}, \bibinfo {author} {\bibfnamefont {P.}~\bibnamefont
  {Magierski}}, \bibinfo {author} {\bibfnamefont {J.~E.}\ \bibnamefont {Drut}},
  \bibinfo {author} {\bibfnamefont {A.}~\bibnamefont {Bulgac}},\ and\ \bibinfo
  {author} {\bibfnamefont {K.~J.}\ \bibnamefont {Roche}},\ }\bibfield  {title}
  {\bibinfo {title} {Cooper pairing above the critical temperature in a unitary
  fermi gas},\ }\href {https://link.aps.org/doi/10.1103/PhysRevLett.110.090401}
  {\bibfield  {journal} {\bibinfo  {journal} {Phys. Rev. Lett.}\ }\textbf
  {\bibinfo {volume} {110}},\ \bibinfo {pages} {090401} (\bibinfo {year}
  {2013})}\BibitemShut {NoStop}%
\bibitem [{\citenamefont {Tajima}\ \emph {et~al.}(2014)\citenamefont {Tajima},
  \citenamefont {Kashimura}, \citenamefont {Hanai}, \citenamefont {Watanabe},\
  and\ \citenamefont {Ohashi}}]{Tajima2014}%
  \BibitemOpen
  \bibfield  {author} {\bibinfo {author} {\bibfnamefont {H.}~\bibnamefont
  {Tajima}}, \bibinfo {author} {\bibfnamefont {T.}~\bibnamefont {Kashimura}},
  \bibinfo {author} {\bibfnamefont {R.}~\bibnamefont {Hanai}}, \bibinfo
  {author} {\bibfnamefont {R.}~\bibnamefont {Watanabe}},\ and\ \bibinfo
  {author} {\bibfnamefont {Y.}~\bibnamefont {Ohashi}},\ }\bibfield  {title}
  {\bibinfo {title} {Uniform spin susceptibility and spin-gap phenomenon in the
  bcs-bec-crossover regime of an ultracold fermi gas},\ }\href
  {https://link.aps.org/doi/10.1103/PhysRevA.89.033617} {\bibfield  {journal}
  {\bibinfo  {journal} {Phys. Rev. A}\ }\textbf {\bibinfo {volume} {89}},\
  \bibinfo {pages} {033617} (\bibinfo {year} {2014})}\BibitemShut {NoStop}%
\bibitem [{\citenamefont {Pantel}\ \emph {et~al.}(2014)\citenamefont {Pantel},
  \citenamefont {Davesne},\ and\ \citenamefont {Urban}}]{Pantel2014}%
  \BibitemOpen
  \bibfield  {author} {\bibinfo {author} {\bibfnamefont {P.-A.}\ \bibnamefont
  {Pantel}}, \bibinfo {author} {\bibfnamefont {D.}~\bibnamefont {Davesne}},\
  and\ \bibinfo {author} {\bibfnamefont {M.}~\bibnamefont {Urban}},\ }\bibfield
   {title} {\bibinfo {title} {Polarized fermi gases at finite temperature in
  the bcs-bec crossover},\ }\href
  {https://link.aps.org/doi/10.1103/PhysRevA.90.053629} {\bibfield  {journal}
  {\bibinfo  {journal} {Phys. Rev. A}\ }\textbf {\bibinfo {volume} {90}},\
  \bibinfo {pages} {053629} (\bibinfo {year} {2014})}\BibitemShut {NoStop}%
\bibitem [{\citenamefont {Chen}\ and\ \citenamefont {Wang}(2014)}]{Chen2014}%
  \BibitemOpen
  \bibfield  {author} {\bibinfo {author} {\bibfnamefont {Q.}~\bibnamefont
  {Chen}}\ and\ \bibinfo {author} {\bibfnamefont {J.}~\bibnamefont {Wang}},\
  }\bibfield  {title} {\bibinfo {title} {Pseudogap phenomena in ultracold
  atomic fermi gases},\ }\href {https://doi.org/10.1007/s11467-014-0448-7}
  {\bibfield  {journal} {\bibinfo  {journal} {Front. Phys.}\ }\textbf {\bibinfo
  {volume} {9}},\ \bibinfo {pages} {539} (\bibinfo {year} {2014})}\BibitemShut
  {NoStop}%
\bibitem [{\citenamefont {Boettcher}\ \emph {et~al.}(2015)\citenamefont
  {Boettcher}, \citenamefont {Braun}, \citenamefont {Herbst}, \citenamefont
  {Pawlowski}, \citenamefont {Roscher},\ and\ \citenamefont
  {Wetterich}}]{Boettcher2014}%
  \BibitemOpen
  \bibfield  {author} {\bibinfo {author} {\bibfnamefont {I.}~\bibnamefont
  {Boettcher}}, \bibinfo {author} {\bibfnamefont {J.}~\bibnamefont {Braun}},
  \bibinfo {author} {\bibfnamefont {T.~K.}\ \bibnamefont {Herbst}}, \bibinfo
  {author} {\bibfnamefont {J.~M.}\ \bibnamefont {Pawlowski}}, \bibinfo {author}
  {\bibfnamefont {D.}~\bibnamefont {Roscher}},\ and\ \bibinfo {author}
  {\bibfnamefont {C.}~\bibnamefont {Wetterich}},\ }\bibfield  {title} {\bibinfo
  {title} {Phase structure of spin-imbalanced unitary fermi gases},\ }\href
  {https://link.aps.org/doi/10.1103/PhysRevA.91.013610} {\bibfield  {journal}
  {\bibinfo  {journal} {Phys. Rev. A}\ }\textbf {\bibinfo {volume} {91}},\
  \bibinfo {pages} {013610} (\bibinfo {year} {2015})}\BibitemShut {NoStop}%
\bibitem [{\citenamefont {Roscher}\ \emph {et~al.}(2015)\citenamefont
  {Roscher}, \citenamefont {Braun},\ and\ \citenamefont {Drut}}]{Roscher2015}%
  \BibitemOpen
  \bibfield  {author} {\bibinfo {author} {\bibfnamefont {D.}~\bibnamefont
  {Roscher}}, \bibinfo {author} {\bibfnamefont {J.}~\bibnamefont {Braun}},\
  and\ \bibinfo {author} {\bibfnamefont {J.~E.}\ \bibnamefont {Drut}},\
  }\bibfield  {title} {\bibinfo {title} {Phase structure of mass- and
  spin-imbalanced unitary fermi gases},\ }\href
  {https://doi.org/10.1103/PhysRevA.91.053611} {\bibfield  {journal} {\bibinfo
  {journal} {Phys. Rev. A}\ }\textbf {\bibinfo {volume} {91}},\ \bibinfo
  {pages} {053611} (\bibinfo {year} {2015})}\BibitemShut {NoStop}%
\bibitem [{\citenamefont {Krippa}(2015)}]{Krippa2015}%
  \BibitemOpen
  \bibfield  {author} {\bibinfo {author} {\bibfnamefont {B.}~\bibnamefont
  {Krippa}},\ }\bibfield  {title} {\bibinfo {title} {Pairing in asymmetric
  many-fermion systems: Functional renormalisation group approach},\ }\href
  {https://doi.org/10.1016/j.physletb.2015.03.057} {\bibfield  {journal}
  {\bibinfo  {journal} {Phys. Lett. B}\ }\textbf {\bibinfo {volume} {744}},\
  \bibinfo {pages} {288–292} (\bibinfo {year} {2015})}\BibitemShut {NoStop}%
\bibitem [{\citenamefont {Jensen}\ \emph
  {et~al.}(2020{\natexlab{a}})\citenamefont {Jensen}, \citenamefont
  {Gilbreth},\ and\ \citenamefont {Alhassid}}]{Jensen2020}%
  \BibitemOpen
  \bibfield  {author} {\bibinfo {author} {\bibfnamefont {S.}~\bibnamefont
  {Jensen}}, \bibinfo {author} {\bibfnamefont {C.~N.}\ \bibnamefont
  {Gilbreth}},\ and\ \bibinfo {author} {\bibfnamefont {Y.}~\bibnamefont
  {Alhassid}},\ }\bibfield  {title} {\bibinfo {title} {Pairing correlations
  across the superfluid phase transition in the unitary fermi gas},\ }\href
  {https://doi.org/10.1103/PhysRevLett.124.090604} {\bibfield  {journal}
  {\bibinfo  {journal} {Phys. Rev. Lett.}\ }\textbf {\bibinfo {volume} {124}},\
  \bibinfo {pages} {090604} (\bibinfo {year} {2020}{\natexlab{a}})}\BibitemShut
  {NoStop}%
\bibitem [{\citenamefont {Jensen}\ \emph {et~al.}(2019)\citenamefont {Jensen},
  \citenamefont {Gilbreth},\ and\ \citenamefont {Alhassid}}]{Jensen2019}%
  \BibitemOpen
  \bibfield  {author} {\bibinfo {author} {\bibfnamefont {S.}~\bibnamefont
  {Jensen}}, \bibinfo {author} {\bibfnamefont {C.~N.}\ \bibnamefont
  {Gilbreth}},\ and\ \bibinfo {author} {\bibfnamefont {Y.}~\bibnamefont
  {Alhassid}},\ }\bibfield  {title} {\bibinfo {title} {The pseudogap regime in
  the unitary fermi gas},\ }\href
  {https://doi.org/10.1140/epjst/e2019-800105-y} {\bibfield  {journal}
  {\bibinfo  {journal} {EPJ ST}\ }\textbf {\bibinfo {volume} {227}},\ \bibinfo
  {pages} {2241} (\bibinfo {year} {2019})}\BibitemShut {NoStop}%
\bibitem [{\citenamefont {Pini}\ \emph {et~al.}(2019)\citenamefont {Pini},
  \citenamefont {Pieri},\ and\ \citenamefont {Strinati}}]{Pini2019}%
  \BibitemOpen
  \bibfield  {author} {\bibinfo {author} {\bibfnamefont {M.}~\bibnamefont
  {Pini}}, \bibinfo {author} {\bibfnamefont {P.}~\bibnamefont {Pieri}},\ and\
  \bibinfo {author} {\bibfnamefont {G.~C.}\ \bibnamefont {Strinati}},\
  }\bibfield  {title} {\bibinfo {title} {Fermi gas throughout the bcs-bec
  crossover: Comparative study of $t$-matrix approaches with various degrees of
  self-consistency},\ }\href {https://doi.org/10.1103/PhysRevB.99.094502}
  {\bibfield  {journal} {\bibinfo  {journal} {Phys. Rev. B}\ }\textbf {\bibinfo
  {volume} {99}},\ \bibinfo {pages} {094502} (\bibinfo {year}
  {2019})}\BibitemShut {NoStop}%
\bibitem [{\citenamefont {Richie-Halford}\ \emph {et~al.}(2020)\citenamefont
  {Richie-Halford}, \citenamefont {Drut},\ and\ \citenamefont
  {Bulgac}}]{Richie-Halford2020}%
  \BibitemOpen
  \bibfield  {author} {\bibinfo {author} {\bibfnamefont {A.}~\bibnamefont
  {Richie-Halford}}, \bibinfo {author} {\bibfnamefont {J.~E.}\ \bibnamefont
  {Drut}},\ and\ \bibinfo {author} {\bibfnamefont {A.}~\bibnamefont {Bulgac}},\
  }\bibfield  {title} {\bibinfo {title} {Emergence of a pseudogap in the
  bcs-bec crossover},\ }\href {https://doi.org/10.1103/PhysRevLett.125.060403}
  {\bibfield  {journal} {\bibinfo  {journal} {Phys. Rev. Lett.}\ }\textbf
  {\bibinfo {volume} {125}},\ \bibinfo {pages} {060403} (\bibinfo {year}
  {2020})}\BibitemShut {NoStop}%
\bibitem [{\citenamefont {Schunck}\ \emph {et~al.}(2007)\citenamefont
  {Schunck}, \citenamefont {Shin}, \citenamefont {Schirotzek}, \citenamefont
  {Zwierlein},\ and\ \citenamefont {Ketterle}}]{Schunck2008}%
  \BibitemOpen
  \bibfield  {author} {\bibinfo {author} {\bibfnamefont {C.~H.}\ \bibnamefont
  {Schunck}}, \bibinfo {author} {\bibfnamefont {Y.}~\bibnamefont {Shin}},
  \bibinfo {author} {\bibfnamefont {A.}~\bibnamefont {Schirotzek}}, \bibinfo
  {author} {\bibfnamefont {M.~W.}\ \bibnamefont {Zwierlein}},\ and\ \bibinfo
  {author} {\bibfnamefont {W.}~\bibnamefont {Ketterle}},\ }\bibfield  {title}
  {\bibinfo {title} {Pairing without superfluidity: The ground state of an
  imbalanced fermi mixture},\ }\href {https://doi.org/10.1126/science.1140749}
  {\bibfield  {journal} {\bibinfo  {journal} {Science}\ }\textbf {\bibinfo
  {volume} {316}},\ \bibinfo {pages} {867} (\bibinfo {year}
  {2007})}\BibitemShut {NoStop}%
\bibitem [{\citenamefont {Shin}\ \emph {et~al.}(2008)\citenamefont {Shin},
  \citenamefont {Schunck}, \citenamefont {Schirotzek},\ and\ \citenamefont
  {Ketterle}}]{Shin2008}%
  \BibitemOpen
  \bibfield  {author} {\bibinfo {author} {\bibfnamefont {Y.-i.}\ \bibnamefont
  {Shin}}, \bibinfo {author} {\bibfnamefont {C.~H.}\ \bibnamefont {Schunck}},
  \bibinfo {author} {\bibfnamefont {A.}~\bibnamefont {Schirotzek}},\ and\
  \bibinfo {author} {\bibfnamefont {W.}~\bibnamefont {Ketterle}},\ }\bibfield
  {title} {\bibinfo {title} {Phase diagram of a two-component fermi gas with
  resonant interactions},\ }\href {https://doi.org/10.1038/nature06473}
  {\bibfield  {journal} {\bibinfo  {journal} {Nature}\ }\textbf {\bibinfo
  {volume} {451}},\ \bibinfo {pages} {689} (\bibinfo {year}
  {2008})}\BibitemShut {NoStop}%
\bibitem [{\citenamefont {Nascimbene}\ \emph {et~al.}(2010)\citenamefont
  {Nascimbene}, \citenamefont {Navon}, \citenamefont {Jiang}, \citenamefont
  {Chevy},\ and\ \citenamefont {Salomon}}]{Nascimbene2010}%
  \BibitemOpen
  \bibfield  {author} {\bibinfo {author} {\bibfnamefont {S.}~\bibnamefont
  {Nascimbene}}, \bibinfo {author} {\bibfnamefont {N.}~\bibnamefont {Navon}},
  \bibinfo {author} {\bibfnamefont {K.~J.}\ \bibnamefont {Jiang}}, \bibinfo
  {author} {\bibfnamefont {F.}~\bibnamefont {Chevy}},\ and\ \bibinfo {author}
  {\bibfnamefont {C.}~\bibnamefont {Salomon}},\ }\bibfield  {title} {\bibinfo
  {title} {Exploring the thermodynamics of a universal fermi gas},\ }\href
  {https://doi.org/10.1038/nature08814} {\bibfield  {journal} {\bibinfo
  {journal} {Nature}\ }\textbf {\bibinfo {volume} {463}},\ \bibinfo {pages}
  {1057} (\bibinfo {year} {2010})}\BibitemShut {NoStop}%
\bibitem [{\citenamefont {Nascimb{\`{e}}ne}\ \emph {et~al.}(2010)\citenamefont
  {Nascimb{\`{e}}ne}, \citenamefont {Navon}, \citenamefont {Chevy},\ and\
  \citenamefont {Salomon}}]{Nascimbene2010b}%
  \BibitemOpen
  \bibfield  {author} {\bibinfo {author} {\bibfnamefont {S.}~\bibnamefont
  {Nascimb{\`{e}}ne}}, \bibinfo {author} {\bibfnamefont {N.}~\bibnamefont
  {Navon}}, \bibinfo {author} {\bibfnamefont {F.}~\bibnamefont {Chevy}},\ and\
  \bibinfo {author} {\bibfnamefont {C.}~\bibnamefont {Salomon}},\ }\bibfield
  {title} {\bibinfo {title} {The equation of state of ultracold bose and fermi
  gases: a few examples},\ }\href
  {https://doi.org/10.1088/1367-2630/12/10/103026} {\bibfield  {journal}
  {\bibinfo  {journal} {New J. Phys.}\ }\textbf {\bibinfo {volume} {12}},\
  \bibinfo {pages} {103026} (\bibinfo {year} {2010})}\BibitemShut {NoStop}%
\bibitem [{\citenamefont {Sommer}\ \emph {et~al.}(2011)\citenamefont {Sommer},
  \citenamefont {Ku}, \citenamefont {Roati},\ and\ \citenamefont
  {Zwierlein}}]{Sommer2011}%
  \BibitemOpen
  \bibfield  {author} {\bibinfo {author} {\bibfnamefont {A.}~\bibnamefont
  {Sommer}}, \bibinfo {author} {\bibfnamefont {M.}~\bibnamefont {Ku}}, \bibinfo
  {author} {\bibfnamefont {G.}~\bibnamefont {Roati}},\ and\ \bibinfo {author}
  {\bibfnamefont {M.~W.}\ \bibnamefont {Zwierlein}},\ }\bibfield  {title}
  {\bibinfo {title} {Universal spin transport in a strongly interacting fermi
  gas},\ }\href {https://doi.org/10.1038/nature09989} {\bibfield  {journal}
  {\bibinfo  {journal} {Nature}\ }\textbf {\bibinfo {volume} {472}},\ \bibinfo
  {pages} {1476} (\bibinfo {year} {2011})}\BibitemShut {NoStop}%
\bibitem [{\citenamefont {Nascimb\`ene}\ \emph {et~al.}(2011)\citenamefont
  {Nascimb\`ene}, \citenamefont {Navon}, \citenamefont {Pilati}, \citenamefont
  {Chevy}, \citenamefont {Giorgini}, \citenamefont {Georges},\ and\
  \citenamefont {Salomon}}]{Nascimbene2011}%
  \BibitemOpen
  \bibfield  {author} {\bibinfo {author} {\bibfnamefont {S.}~\bibnamefont
  {Nascimb\`ene}}, \bibinfo {author} {\bibfnamefont {N.}~\bibnamefont {Navon}},
  \bibinfo {author} {\bibfnamefont {S.}~\bibnamefont {Pilati}}, \bibinfo
  {author} {\bibfnamefont {F.}~\bibnamefont {Chevy}}, \bibinfo {author}
  {\bibfnamefont {S.}~\bibnamefont {Giorgini}}, \bibinfo {author}
  {\bibfnamefont {A.}~\bibnamefont {Georges}},\ and\ \bibinfo {author}
  {\bibfnamefont {C.}~\bibnamefont {Salomon}},\ }\bibfield  {title} {\bibinfo
  {title} {Fermi-liquid behavior of the normal phase of a strongly interacting
  gas of cold atoms},\ }\href
  {https://link.aps.org/doi/10.1103/PhysRevLett.106.215303} {\bibfield
  {journal} {\bibinfo  {journal} {Phys. Rev. Lett.}\ }\textbf {\bibinfo
  {volume} {106}},\ \bibinfo {pages} {215303} (\bibinfo {year}
  {2011})}\BibitemShut {NoStop}%
\bibitem [{\citenamefont {Ku}\ \emph {et~al.}(2012)\citenamefont {Ku},
  \citenamefont {Sommer}, \citenamefont {Cheuk},\ and\ \citenamefont
  {Zwierlein}}]{Ku2012}%
  \BibitemOpen
  \bibfield  {author} {\bibinfo {author} {\bibfnamefont {M.~J.~H.}\
  \bibnamefont {Ku}}, \bibinfo {author} {\bibfnamefont {A.~T.}\ \bibnamefont
  {Sommer}}, \bibinfo {author} {\bibfnamefont {L.~W.}\ \bibnamefont {Cheuk}},\
  and\ \bibinfo {author} {\bibfnamefont {M.~W.}\ \bibnamefont {Zwierlein}},\
  }\bibfield  {title} {\bibinfo {title} {Revealing the superfluid lambda
  transition in the universal thermodynamics of a unitary fermi gas},\ }\href
  {https://doi.org/10.1126/science.1214987} {\bibfield  {journal} {\bibinfo
  {journal} {Science}\ }\textbf {\bibinfo {volume} {335}},\ \bibinfo {pages}
  {563} (\bibinfo {year} {2012})}\BibitemShut {NoStop}%
\bibitem [{\citenamefont {Sagi}\ \emph {et~al.}(2015)\citenamefont {Sagi},
  \citenamefont {Drake}, \citenamefont {Paudel}, \citenamefont {Chapurin},\
  and\ \citenamefont {Jin}}]{Sagi2015}%
  \BibitemOpen
  \bibfield  {author} {\bibinfo {author} {\bibfnamefont {Y.}~\bibnamefont
  {Sagi}}, \bibinfo {author} {\bibfnamefont {T.~E.}\ \bibnamefont {Drake}},
  \bibinfo {author} {\bibfnamefont {R.}~\bibnamefont {Paudel}}, \bibinfo
  {author} {\bibfnamefont {R.}~\bibnamefont {Chapurin}},\ and\ \bibinfo
  {author} {\bibfnamefont {D.~S.}\ \bibnamefont {Jin}},\ }\bibfield  {title}
  {\bibinfo {title} {Breakdown of the fermi liquid description for strongly
  interacting fermions},\ }\href
  {https://link.aps.org/doi/10.1103/PhysRevLett.114.075301} {\bibfield
  {journal} {\bibinfo  {journal} {Phys. Rev. Lett.}\ }\textbf {\bibinfo
  {volume} {114}},\ \bibinfo {pages} {075301} (\bibinfo {year}
  {2015})}\BibitemShut {NoStop}%
\bibitem [{\citenamefont {Horikoshi}\ \emph {et~al.}(2017)\citenamefont
  {Horikoshi}, \citenamefont {Koashi}, \citenamefont {Tajima}, \citenamefont
  {Ohashi},\ and\ \citenamefont {Kuwata-Gonokami}}]{Horikoshi2017}%
  \BibitemOpen
  \bibfield  {author} {\bibinfo {author} {\bibfnamefont {M.}~\bibnamefont
  {Horikoshi}}, \bibinfo {author} {\bibfnamefont {M.}~\bibnamefont {Koashi}},
  \bibinfo {author} {\bibfnamefont {H.}~\bibnamefont {Tajima}}, \bibinfo
  {author} {\bibfnamefont {Y.}~\bibnamefont {Ohashi}},\ and\ \bibinfo {author}
  {\bibfnamefont {M.}~\bibnamefont {Kuwata-Gonokami}},\ }\bibfield  {title}
  {\bibinfo {title} {Ground-state thermodynamic quantities of homogeneous
  spin-$1/2$ fermions from the bcs region to the unitarity limit},\ }\href
  {https://doi.org/10.1103/PhysRevX.7.041004} {\bibfield  {journal} {\bibinfo
  {journal} {Phys. Rev. X}\ }\textbf {\bibinfo {volume} {7}},\ \bibinfo {pages}
  {041004} (\bibinfo {year} {2017})}\BibitemShut {NoStop}%
\bibitem [{\citenamefont {Murthy}\ \emph {et~al.}(2018)\citenamefont {Murthy},
  \citenamefont {Neidig}, \citenamefont {Klemt}, \citenamefont {Bayha},
  \citenamefont {Boettcher}, \citenamefont {Enss}, \citenamefont {Holten},
  \citenamefont {Z{\"u}rn}, \citenamefont {Preiss},\ and\ \citenamefont
  {Jochim}}]{Murthy2018}%
  \BibitemOpen
  \bibfield  {author} {\bibinfo {author} {\bibfnamefont {P.~A.}\ \bibnamefont
  {Murthy}}, \bibinfo {author} {\bibfnamefont {M.}~\bibnamefont {Neidig}},
  \bibinfo {author} {\bibfnamefont {R.}~\bibnamefont {Klemt}}, \bibinfo
  {author} {\bibfnamefont {L.}~\bibnamefont {Bayha}}, \bibinfo {author}
  {\bibfnamefont {I.}~\bibnamefont {Boettcher}}, \bibinfo {author}
  {\bibfnamefont {T.}~\bibnamefont {Enss}}, \bibinfo {author} {\bibfnamefont
  {M.}~\bibnamefont {Holten}}, \bibinfo {author} {\bibfnamefont
  {G.}~\bibnamefont {Z{\"u}rn}}, \bibinfo {author} {\bibfnamefont {P.~M.}\
  \bibnamefont {Preiss}},\ and\ \bibinfo {author} {\bibfnamefont
  {S.}~\bibnamefont {Jochim}},\ }\bibfield  {title} {\bibinfo {title}
  {High-temperature pairing in a strongly interacting two-dimensional fermi
  gas},\ }\href {https://doi.org/10.1126/science.aan5950} {\bibfield  {journal}
  {\bibinfo  {journal} {Science}\ }\textbf {\bibinfo {volume} {359}},\ \bibinfo
  {pages} {452} (\bibinfo {year} {2018})}\BibitemShut {NoStop}%
\bibitem [{\citenamefont {Jensen}\ \emph
  {et~al.}(2020{\natexlab{b}})\citenamefont {Jensen}, \citenamefont
  {Gilbreth},\ and\ \citenamefont {Alhassid}}]{Jensen2019b}%
  \BibitemOpen
  \bibfield  {author} {\bibinfo {author} {\bibfnamefont {S.}~\bibnamefont
  {Jensen}}, \bibinfo {author} {\bibfnamefont {C.~N.}\ \bibnamefont
  {Gilbreth}},\ and\ \bibinfo {author} {\bibfnamefont {Y.}~\bibnamefont
  {Alhassid}},\ }\bibfield  {title} {\bibinfo {title} {Contact in the unitary
  fermi gas across the superfluid phase transition},\ }\href
  {https://doi.org/10.1103/PhysRevLett.125.043402} {\bibfield  {journal}
  {\bibinfo  {journal} {Phys. Rev. Lett.}\ }\textbf {\bibinfo {volume} {125}},\
  \bibinfo {pages} {043402} (\bibinfo {year} {2020}{\natexlab{b}})}\BibitemShut
  {NoStop}%
\bibitem [{\citenamefont {Long}\ \emph {et~al.}(2021)\citenamefont {Long},
  \citenamefont {Xiong},\ and\ \citenamefont {Parker}}]{Long2020}%
  \BibitemOpen
  \bibfield  {author} {\bibinfo {author} {\bibfnamefont {Y.}~\bibnamefont
  {Long}}, \bibinfo {author} {\bibfnamefont {F.}~\bibnamefont {Xiong}},\ and\
  \bibinfo {author} {\bibfnamefont {C.~V.}\ \bibnamefont {Parker}},\ }\bibfield
   {title} {\bibinfo {title} {Spin susceptibility above the superfluid onset in
  ultracold fermi gases},\ }\href
  {https://doi.org/10.1103/PhysRevLett.126.153402} {\bibfield  {journal}
  {\bibinfo  {journal} {Phys. Rev. Lett.}\ }\textbf {\bibinfo {volume} {126}},\
  \bibinfo {pages} {153402} (\bibinfo {year} {2021})}\BibitemShut {NoStop}%
\bibitem [{\citenamefont {Zwerger~(Ed.)}(2012)}]{Zwerger2012}%
  \BibitemOpen
  \bibfield  {author} {\bibinfo {author} {\bibfnamefont {W.}~\bibnamefont
  {Zwerger~(Ed.)}},\ }\href {https://doi.org/10.1007/978-3-642-21978-8} {\emph
  {\bibinfo {title} {The BCS-BEC Crossover and the Unitary Fermi Gas}}}\
  (\bibinfo  {publisher} {Springer-Verlag},\ \bibinfo {address} {Berlin
  Heidelberg},\ \bibinfo {year} {2012})\BibitemShut {NoStop}%
\bibitem [{\citenamefont {Nishida}\ and\ \citenamefont
  {Son}(2007{\natexlab{a}})}]{Nishida2007}%
  \BibitemOpen
  \bibfield  {author} {\bibinfo {author} {\bibfnamefont {Y.}~\bibnamefont
  {Nishida}}\ and\ \bibinfo {author} {\bibfnamefont {D.~T.}\ \bibnamefont
  {Son}},\ }\bibfield  {title} {\bibinfo {title} {Nonrelativistic conformal
  field theories},\ }\href {https://doi.org/10.1103/PhysRevD.76.086004}
  {\bibfield  {journal} {\bibinfo  {journal} {Phys. Rev. D}\ }\textbf {\bibinfo
  {volume} {76}},\ \bibinfo {pages} {086004} (\bibinfo {year}
  {2007}{\natexlab{a}})}\BibitemShut {NoStop}%
\bibitem [{\citenamefont {Ho}(2004)}]{Ho2004}%
  \BibitemOpen
  \bibfield  {author} {\bibinfo {author} {\bibfnamefont {T.-L.}\ \bibnamefont
  {Ho}},\ }\bibfield  {title} {\bibinfo {title} {Universal thermodynamics of
  degenerate quantum gases in the unitarity limit},\ }\href
  {https://doi.org/10.1103/PhysRevLett.92.090402} {\bibfield  {journal}
  {\bibinfo  {journal} {Phys. Rev. Lett.}\ }\textbf {\bibinfo {volume} {92}},\
  \bibinfo {pages} {090402} (\bibinfo {year} {2004})}\BibitemShut {NoStop}%
\bibitem [{\citenamefont {Braaten}\ and\ \citenamefont
  {Hammer}(2006)}]{Braaten2006}%
  \BibitemOpen
  \bibfield  {author} {\bibinfo {author} {\bibfnamefont {E.}~\bibnamefont
  {Braaten}}\ and\ \bibinfo {author} {\bibfnamefont {H.-W.}\ \bibnamefont
  {Hammer}},\ }\bibfield  {title} {\bibinfo {title} {Universality in few-body
  systems with large scattering length},\ }\href
  {https://doi.org/https://doi.org/10.1016/j.physrep.2006.03.001} {\bibfield
  {journal} {\bibinfo  {journal} {Phys. Rep.}\ }\textbf {\bibinfo {volume}
  {428}},\ \bibinfo {pages} {259 } (\bibinfo {year} {2006})}\BibitemShut
  {NoStop}%
\bibitem [{\citenamefont {Inguscio}\ \emph {et~al.}(208)\citenamefont
  {Inguscio}, \citenamefont {Ketterle},\ and\ \citenamefont
  {Salomon}}]{ExpReview}%
  \BibitemOpen
  \bibinfo {editor} {\bibfnamefont {M.}~\bibnamefont {Inguscio}}, \bibinfo
  {editor} {\bibfnamefont {W.}~\bibnamefont {Ketterle}},\ and\ \bibinfo
  {editor} {\bibfnamefont {C.}~\bibnamefont {Salomon}},\ eds.,\ \href@noop {}
  {\emph {\bibinfo {title} {Proceedings of the International School of Physics
  ``Enrico Fermi", Course CLXIV, Varenna}}}\ (\bibinfo  {publisher} {IOS},\
  \bibinfo {address} {Amsterdam},\ \bibinfo {year} {208})\BibitemShut {NoStop}%
\bibitem [{\citenamefont {Gaebler}\ \emph {et~al.}(2010)\citenamefont
  {Gaebler}, \citenamefont {Stewart}, \citenamefont {Drake}, \citenamefont
  {Jin}, \citenamefont {Perali}, \citenamefont {Pieri},\ and\ \citenamefont
  {Strinati}}]{Gaebler2010}%
  \BibitemOpen
  \bibfield  {author} {\bibinfo {author} {\bibfnamefont {J.~P.}\ \bibnamefont
  {Gaebler}}, \bibinfo {author} {\bibfnamefont {J.~T.}\ \bibnamefont
  {Stewart}}, \bibinfo {author} {\bibfnamefont {T.~E.}\ \bibnamefont {Drake}},
  \bibinfo {author} {\bibfnamefont {D.~S.}\ \bibnamefont {Jin}}, \bibinfo
  {author} {\bibfnamefont {A.}~\bibnamefont {Perali}}, \bibinfo {author}
  {\bibfnamefont {P.}~\bibnamefont {Pieri}},\ and\ \bibinfo {author}
  {\bibfnamefont {G.~C.}\ \bibnamefont {Strinati}},\ }\bibfield  {title}
  {\bibinfo {title} {Observation of pseudogap behaviour in a strongly
  interacting fermi gas},\ }\href {https://doi.org/10.1038/nphys1709}
  {\bibfield  {journal} {\bibinfo  {journal} {Nat. Phys.}\ }\textbf {\bibinfo
  {volume} {6}},\ \bibinfo {pages} {569} (\bibinfo {year} {2010})}\BibitemShut
  {NoStop}%
\bibitem [{\citenamefont {Mueller}(2017)}]{Mueller2017}%
  \BibitemOpen
  \bibfield  {author} {\bibinfo {author} {\bibfnamefont {E.~J.}\ \bibnamefont
  {Mueller}},\ }\bibfield  {title} {\bibinfo {title} {Review of pseudogaps in
  strongly interacting fermi gases},\ }\href
  {https://doi.org/10.1088%2F1361-6633%2Faa7e53} {\bibfield  {journal}
  {\bibinfo  {journal} {Rep. Prog. Phys.}\ }\textbf {\bibinfo {volume} {80}},\
  \bibinfo {pages} {104401} (\bibinfo {year} {2017})}\BibitemShut {NoStop}%
\bibitem [{\citenamefont {Rothstein}\ and\ \citenamefont
  {Shrivastava}(2019)}]{Rothstein2019}%
  \BibitemOpen
  \bibfield  {author} {\bibinfo {author} {\bibfnamefont {I.~Z.}\ \bibnamefont
  {Rothstein}}\ and\ \bibinfo {author} {\bibfnamefont {P.}~\bibnamefont
  {Shrivastava}},\ }\bibfield  {title} {\bibinfo {title} {{Symmetry obstruction
  to Fermi liquid behavior in the unitary limit}},\ }\href
  {https://doi.org/10.1103/PhysRevB.99.035101} {\bibfield  {journal} {\bibinfo
  {journal} {Phys. Rev. B}\ }\textbf {\bibinfo {volume} {99}},\ \bibinfo
  {pages} {035101} (\bibinfo {year} {2019})}\BibitemShut {NoStop}%
\bibitem [{\citenamefont {Pieri}\ \emph {et~al.}(2004)\citenamefont {Pieri},
  \citenamefont {Pisani},\ and\ \citenamefont {Strinati}}]{Pieri2004}%
  \BibitemOpen
  \bibfield  {author} {\bibinfo {author} {\bibfnamefont {P.}~\bibnamefont
  {Pieri}}, \bibinfo {author} {\bibfnamefont {L.}~\bibnamefont {Pisani}},\ and\
  \bibinfo {author} {\bibfnamefont {G.~C.}\ \bibnamefont {Strinati}},\
  }\bibfield  {title} {\bibinfo {title} {Bcs-bec crossover at finite
  temperature in the broken-symmetry phase},\ }\href
  {https://doi.org/10.1103/PhysRevB.70.094508} {\bibfield  {journal} {\bibinfo
  {journal} {Phys. Rev. B}\ }\textbf {\bibinfo {volume} {70}},\ \bibinfo
  {pages} {094508} (\bibinfo {year} {2004})}\BibitemShut {NoStop}%
\bibitem [{\citenamefont {Haussmann}\ \emph {et~al.}(2009)\citenamefont
  {Haussmann}, \citenamefont {Punk},\ and\ \citenamefont
  {Zwerger}}]{Haussmann2009}%
  \BibitemOpen
  \bibfield  {author} {\bibinfo {author} {\bibfnamefont {R.}~\bibnamefont
  {Haussmann}}, \bibinfo {author} {\bibfnamefont {M.}~\bibnamefont {Punk}},\
  and\ \bibinfo {author} {\bibfnamefont {W.}~\bibnamefont {Zwerger}},\
  }\bibfield  {title} {\bibinfo {title} {Spectral functions and rf response of
  ultracold fermionic atoms},\ }\href
  {https://doi.org/10.1103/PhysRevA.80.063612} {\bibfield  {journal} {\bibinfo
  {journal} {Phys. Rev. A}\ }\textbf {\bibinfo {volume} {80}},\ \bibinfo
  {pages} {063612} (\bibinfo {year} {2009})}\BibitemShut {NoStop}%
\bibitem [{\citenamefont {Trivedi}\ and\ \citenamefont
  {Randeria}(1995)}]{Trivedi1995}%
  \BibitemOpen
  \bibfield  {author} {\bibinfo {author} {\bibfnamefont {N.}~\bibnamefont
  {Trivedi}}\ and\ \bibinfo {author} {\bibfnamefont {M.}~\bibnamefont
  {Randeria}},\ }\bibfield  {title} {\bibinfo {title} {Deviations from
  fermi-liquid behavior above ${T}_{c}$ in 2d short coherence length
  superconductors},\ }\href
  {https://link.aps.org/doi/10.1103/PhysRevLett.75.312} {\bibfield  {journal}
  {\bibinfo  {journal} {Phys. Rev. Lett.}\ }\textbf {\bibinfo {volume} {75}},\
  \bibinfo {pages} {312} (\bibinfo {year} {1995})}\BibitemShut {NoStop}%
\bibitem [{\citenamefont {Randeria}\ \emph {et~al.}(1992)\citenamefont
  {Randeria}, \citenamefont {Trivedi}, \citenamefont {Moreo},\ and\
  \citenamefont {Scalettar}}]{Randeria1992}%
  \BibitemOpen
  \bibfield  {author} {\bibinfo {author} {\bibfnamefont {M.}~\bibnamefont
  {Randeria}}, \bibinfo {author} {\bibfnamefont {N.}~\bibnamefont {Trivedi}},
  \bibinfo {author} {\bibfnamefont {A.}~\bibnamefont {Moreo}},\ and\ \bibinfo
  {author} {\bibfnamefont {R.~T.}\ \bibnamefont {Scalettar}},\ }\bibfield
  {title} {\bibinfo {title} {Pairing and spin gap in the normal state of short
  coherence length superconductors},\ }\href
  {https://link.aps.org/doi/10.1103/PhysRevLett.69.2001} {\bibfield  {journal}
  {\bibinfo  {journal} {Phys. Rev. Lett.}\ }\textbf {\bibinfo {volume} {69}},\
  \bibinfo {pages} {2001} (\bibinfo {year} {1992})}\BibitemShut {NoStop}%
\bibitem [{\citenamefont {Berger}\ \emph {et~al.}(2021)\citenamefont {Berger},
  \citenamefont {Rammelmüller}, \citenamefont {Loheac}, \citenamefont
  {Ehmann}, \citenamefont {Braun},\ and\ \citenamefont {Drut}}]{Berger2021}%
  \BibitemOpen
  \bibfield  {author} {\bibinfo {author} {\bibfnamefont {C.}~\bibnamefont
  {Berger}}, \bibinfo {author} {\bibfnamefont {L.}~\bibnamefont
  {Rammelmüller}}, \bibinfo {author} {\bibfnamefont {A.}~\bibnamefont
  {Loheac}}, \bibinfo {author} {\bibfnamefont {F.}~\bibnamefont {Ehmann}},
  \bibinfo {author} {\bibfnamefont {J.}~\bibnamefont {Braun}},\ and\ \bibinfo
  {author} {\bibfnamefont {J.}~\bibnamefont {Drut}},\ }\bibfield  {title}
  {\bibinfo {title} {Complex langevin and other approaches to the sign problem
  in quantum many-body physics},\ }\href
  {https://doi.org/https://doi.org/10.1016/j.physrep.2020.09.002} {\bibfield
  {journal} {\bibinfo  {journal} {Phys. Rep.}\ }\textbf {\bibinfo {volume}
  {892}},\ \bibinfo {pages} {1} (\bibinfo {year} {2021})}\BibitemShut {NoStop}%
\bibitem [{\citenamefont {Attanasio}\ \emph {et~al.}(2020)\citenamefont
  {Attanasio}, \citenamefont {J{\"a}ger},\ and\ \citenamefont
  {Ziegler}}]{Attanasio2020}%
  \BibitemOpen
  \bibfield  {author} {\bibinfo {author} {\bibfnamefont {F.}~\bibnamefont
  {Attanasio}}, \bibinfo {author} {\bibfnamefont {B.}~\bibnamefont
  {J{\"a}ger}},\ and\ \bibinfo {author} {\bibfnamefont {F.~P.~G.}\ \bibnamefont
  {Ziegler}},\ }\bibfield  {title} {\bibinfo {title} {Complex langevin
  simulations and the qcd phase diagram: recent developments},\ }\href
  {https://doi.org/10.1140/epja/s10050-020-00256-z} {\bibfield  {journal}
  {\bibinfo  {journal} {Eur. Phys. J. A}\ }\textbf {\bibinfo {volume} {56}},\
  \bibinfo {pages} {251} (\bibinfo {year} {2020})}\BibitemShut {NoStop}%
\bibitem [{\citenamefont {Hou}\ and\ \citenamefont
  {Drut}(2020{\natexlab{a}})}]{Hou2020b}%
  \BibitemOpen
  \bibfield  {author} {\bibinfo {author} {\bibfnamefont {Y.}~\bibnamefont
  {Hou}}\ and\ \bibinfo {author} {\bibfnamefont {J.~E.}\ \bibnamefont {Drut}},\
  }\bibfield  {title} {\bibinfo {title} {Fourth- and fifth-order virial
  coefficients from weak coupling to unitarity},\ }\href
  {https://doi.org/10.1103/PhysRevLett.125.050403} {\bibfield  {journal}
  {\bibinfo  {journal} {Phys. Rev. Lett.}\ }\textbf {\bibinfo {volume} {125}},\
  \bibinfo {pages} {050403} (\bibinfo {year} {2020}{\natexlab{a}})}\BibitemShut
  {NoStop}%
\bibitem [{\citenamefont {Hou}\ and\ \citenamefont
  {Drut}(2020{\natexlab{b}})}]{Hou2020}%
  \BibitemOpen
  \bibfield  {author} {\bibinfo {author} {\bibfnamefont {Y.}~\bibnamefont
  {Hou}}\ and\ \bibinfo {author} {\bibfnamefont {J.~E.}\ \bibnamefont {Drut}},\
  }\bibfield  {title} {\bibinfo {title} {Virial expansion of attractively
  interacting fermi gases in one, two, and three dimensions, up to fifth
  order},\ }\href {https://doi.org/10.1103/PhysRevA.102.033319} {\bibfield
  {journal} {\bibinfo  {journal} {Phys. Rev. A}\ }\textbf {\bibinfo {volume}
  {102}},\ \bibinfo {pages} {033319} (\bibinfo {year}
  {2020}{\natexlab{b}})}\BibitemShut {NoStop}%
\bibitem [{\citenamefont {Strinati}\ \emph {et~al.}(2018)\citenamefont
  {Strinati}, \citenamefont {Pieri}, \citenamefont {R{\"o}pke}, \citenamefont
  {Schuck},\ and\ \citenamefont {Urban}}]{Strinati2018}%
  \BibitemOpen
  \bibfield  {author} {\bibinfo {author} {\bibfnamefont {G.~C.}\ \bibnamefont
  {Strinati}}, \bibinfo {author} {\bibfnamefont {P.}~\bibnamefont {Pieri}},
  \bibinfo {author} {\bibfnamefont {G.}~\bibnamefont {R{\"o}pke}}, \bibinfo
  {author} {\bibfnamefont {P.}~\bibnamefont {Schuck}},\ and\ \bibinfo {author}
  {\bibfnamefont {M.}~\bibnamefont {Urban}},\ }\bibfield  {title} {\bibinfo
  {title} {{The {BCS}{\textendash}{BEC} crossover: From ultra-cold Fermi gases
  to nuclear systems}},\ }\href {https://doi.org/10.1016/j.physrep.2018.02.004}
  {\bibfield  {journal} {\bibinfo  {journal} {Phys. Rep.}\ }\textbf {\bibinfo
  {volume} {738}},\ \bibinfo {pages} {1} (\bibinfo {year} {2018})}\BibitemShut
  {NoStop}%
\bibitem [{\citenamefont {Loheac}\ and\ \citenamefont
  {Drut}(2017)}]{Loheac2017}%
  \BibitemOpen
  \bibfield  {author} {\bibinfo {author} {\bibfnamefont {A.~C.}\ \bibnamefont
  {Loheac}}\ and\ \bibinfo {author} {\bibfnamefont {J.~E.}\ \bibnamefont
  {Drut}},\ }\bibfield  {title} {\bibinfo {title} {Third-order perturbative
  lattice and complex langevin analyses of the finite-temperature equation of
  state of nonrelativistic fermions in one dimension},\ }\href
  {https://link.aps.org/doi/10.1103/PhysRevD.95.094502} {\bibfield  {journal}
  {\bibinfo  {journal} {Phys. Rev. D}\ }\textbf {\bibinfo {volume} {95}},\
  \bibinfo {pages} {094502} (\bibinfo {year} {2017})}\BibitemShut {NoStop}%
\bibitem [{\citenamefont {Rammelm\"uller}\ \emph {et~al.}(2017)\citenamefont
  {Rammelm\"uller}, \citenamefont {Porter}, \citenamefont {Drut},\ and\
  \citenamefont {Braun}}]{Rammelmueller2017}%
  \BibitemOpen
  \bibfield  {author} {\bibinfo {author} {\bibfnamefont {L.}~\bibnamefont
  {Rammelm\"uller}}, \bibinfo {author} {\bibfnamefont {W.~J.}\ \bibnamefont
  {Porter}}, \bibinfo {author} {\bibfnamefont {J.~E.}\ \bibnamefont {Drut}},\
  and\ \bibinfo {author} {\bibfnamefont {J.}~\bibnamefont {Braun}},\ }\bibfield
   {title} {\bibinfo {title} {Surmounting the sign problem in nonrelativistic
  calculations: A case study with mass-imbalanced fermions},\ }\href
  {https://link.aps.org/doi/10.1103/PhysRevD.96.094506} {\bibfield  {journal}
  {\bibinfo  {journal} {Phys. Rev. D}\ }\textbf {\bibinfo {volume} {96}},\
  \bibinfo {pages} {094506} (\bibinfo {year} {2017})}\BibitemShut {NoStop}%
\bibitem [{\citenamefont {Rammelm\"uller}\ \emph {et~al.}(2018)\citenamefont
  {Rammelm\"uller}, \citenamefont {Loheac}, \citenamefont {Drut},\ and\
  \citenamefont {Braun}}]{Rammelmueller2018}%
  \BibitemOpen
  \bibfield  {author} {\bibinfo {author} {\bibfnamefont {L.}~\bibnamefont
  {Rammelm\"uller}}, \bibinfo {author} {\bibfnamefont {A.~C.}\ \bibnamefont
  {Loheac}}, \bibinfo {author} {\bibfnamefont {J.~E.}\ \bibnamefont {Drut}},\
  and\ \bibinfo {author} {\bibfnamefont {J.}~\bibnamefont {Braun}},\ }\bibfield
   {title} {\bibinfo {title} {Finite-temperature equation of state of polarized
  fermions at unitarity},\ }\href
  {https://doi.org/10.1103/PhysRevLett.121.173001} {\bibfield  {journal}
  {\bibinfo  {journal} {Phys. Rev. Lett.}\ }\textbf {\bibinfo {volume} {121}},\
  \bibinfo {pages} {173001} (\bibinfo {year} {2018})}\BibitemShut {NoStop}%
\bibitem [{\citenamefont {Rammelmüller}\ \emph {et~al.}(2020)\citenamefont
  {Rammelmüller}, \citenamefont {Drut},\ and\ \citenamefont
  {Braun}}]{Rammelmueller2020}%
  \BibitemOpen
  \bibfield  {author} {\bibinfo {author} {\bibfnamefont {L.}~\bibnamefont
  {Rammelmüller}}, \bibinfo {author} {\bibfnamefont {J.~E.}\ \bibnamefont
  {Drut}},\ and\ \bibinfo {author} {\bibfnamefont {J.}~\bibnamefont {Braun}},\
  }\bibfield  {title} {\bibinfo {title} {{Pairing patterns in one-dimensional
  spin- and mass-imbalanced Fermi gases}},\ }\href
  {https://doi.org/10.21468/SciPostPhys.9.1.014} {\bibfield  {journal}
  {\bibinfo  {journal} {SciPost Phys.}\ }\textbf {\bibinfo {volume} {9}},\
  \bibinfo {pages} {14} (\bibinfo {year} {2020})}\BibitemShut {NoStop}%
\bibitem [{\citenamefont {Drut}\ and\ \citenamefont
  {Nicholson}(2013)}]{Drut2013}%
  \BibitemOpen
  \bibfield  {author} {\bibinfo {author} {\bibfnamefont {J.~E.}\ \bibnamefont
  {Drut}}\ and\ \bibinfo {author} {\bibfnamefont {A.~N.}\ \bibnamefont
  {Nicholson}},\ }\bibfield  {title} {\bibinfo {title} {Lattice methods for
  strongly interacting many-body systems},\ }\href
  {http://stacks.iop.org/0954-3899/40/i=4/a=043101} {\bibfield  {journal}
  {\bibinfo  {journal} {J. Phys. G}\ }\textbf {\bibinfo {volume} {40}},\
  \bibinfo {pages} {043101} (\bibinfo {year} {2013})}\BibitemShut {NoStop}%
\bibitem [{\citenamefont {Lee}(2017)}]{Lee2017}%
  \BibitemOpen
  \bibfield  {author} {\bibinfo {author} {\bibfnamefont {D.}~\bibnamefont
  {Lee}},\ }\bibinfo {title} {Lattice methods and the nuclear few- and
  many-body problem},\ in\ \href
  {"https://doi.org/10.1007/978-3-319-53336-0_6"} {\emph {\bibinfo {booktitle}
  {"An Advanced Course in Computational Nuclear Physics: Bridging the Scales
  from Quarks to Neutron Stars"}}},\ \bibinfo {editor} {edited by\ \bibinfo
  {editor} {\bibfnamefont {M.}~\bibnamefont {Hjorth-Jensen}}, \bibinfo {editor}
  {\bibfnamefont {M.~P.}\ \bibnamefont {Lombardo}},\ and\ \bibinfo {editor}
  {\bibfnamefont {U.}~\bibnamefont {van Kolck}}}\ (\bibinfo  {publisher}
  {Springer International Publishing},\ \bibinfo {address} {Cham},\ \bibinfo
  {year} {2017})\ pp.\ \bibinfo {pages} {237--261}\BibitemShut {NoStop}%
\bibitem [{\citenamefont {Lüscher}(1986{\natexlab{a}})}]{Luescher1986a}%
  \BibitemOpen
  \bibfield  {author} {\bibinfo {author} {\bibfnamefont {M.}~\bibnamefont
  {Lüscher}},\ }\bibfield  {title} {\bibinfo {title} {Volume dependence of the
  energy spectrum in massive quantum field theories. i. stable particle
  states},\ }\href {https://projecteuclid.org:443/euclid.cmp/1104114998}
  {\bibfield  {journal} {\bibinfo  {journal} {Comm. Math. Phys.}\ }\textbf
  {\bibinfo {volume} {104}},\ \bibinfo {pages} {177} (\bibinfo {year}
  {1986}{\natexlab{a}})}\BibitemShut {NoStop}%
\bibitem [{\citenamefont {Lüscher}(1986{\natexlab{b}})}]{Luescher1986b}%
  \BibitemOpen
  \bibfield  {author} {\bibinfo {author} {\bibfnamefont {M.}~\bibnamefont
  {Lüscher}},\ }\bibfield  {title} {\bibinfo {title} {Volume dependence of the
  energy spectrum in massive quantum field theories. ii. scattering states},\
  }\href {https://projecteuclid.org:443/euclid.cmp/1104115329} {\bibfield
  {journal} {\bibinfo  {journal} {Comm. Math. Phys.}\ }\textbf {\bibinfo
  {volume} {105}},\ \bibinfo {pages} {153} (\bibinfo {year}
  {1986}{\natexlab{b}})}\BibitemShut {NoStop}%
\bibitem [{\citenamefont {Burovski}\ \emph {et~al.}(2006)\citenamefont
  {Burovski}, \citenamefont {Prokof{\'e}v}, \citenamefont {Svistunov},\ and\
  \citenamefont {Troyer}}]{Burovski2006}%
  \BibitemOpen
  \bibfield  {author} {\bibinfo {author} {\bibfnamefont {E.}~\bibnamefont
  {Burovski}}, \bibinfo {author} {\bibfnamefont {N.}~\bibnamefont
  {Prokof{\'e}v}}, \bibinfo {author} {\bibfnamefont {B.}~\bibnamefont
  {Svistunov}},\ and\ \bibinfo {author} {\bibfnamefont {M.}~\bibnamefont
  {Troyer}},\ }\bibfield  {title} {\bibinfo {title} {The
  fermi{\textendash}hubbard model at unitarity},\ }\href
  {https://doi.org/10.1088%2F1367-2630%2F8%2F8%2F153} {\bibfield  {journal}
  {\bibinfo  {journal} {New J. Phys.}\ }\textbf {\bibinfo {volume} {8}},\
  \bibinfo {pages} {153} (\bibinfo {year} {2006})}\BibitemShut {NoStop}%
\bibitem [{\citenamefont {Bulgac}\ \emph {et~al.}(2008)\citenamefont {Bulgac},
  \citenamefont {Drut},\ and\ \citenamefont {Magierski}}]{Bulgac2008}%
  \BibitemOpen
  \bibfield  {author} {\bibinfo {author} {\bibfnamefont {A.}~\bibnamefont
  {Bulgac}}, \bibinfo {author} {\bibfnamefont {J.~E.}\ \bibnamefont {Drut}},\
  and\ \bibinfo {author} {\bibfnamefont {P.}~\bibnamefont {Magierski}},\
  }\bibfield  {title} {\bibinfo {title} {Quantum monte carlo simulations of the
  bcs-bec crossover at finite temperature},\ }\href
  {https://doi.org/10.1103/PhysRevA.78.023625} {\bibfield  {journal} {\bibinfo
  {journal} {Phys. Rev. A}\ }\textbf {\bibinfo {volume} {78}},\ \bibinfo
  {pages} {023625} (\bibinfo {year} {2008})}\BibitemShut {NoStop}%
\bibitem [{\citenamefont {Lee}(2009)}]{Lee2009}%
  \BibitemOpen
  \bibfield  {author} {\bibinfo {author} {\bibfnamefont {D.}~\bibnamefont
  {Lee}},\ }\bibfield  {title} {\bibinfo {title} {Lattice simulations for few-
  and many-body systems},\ }\href
  {https://doi.org/https://doi.org/10.1016/j.ppnp.2008.12.001} {\bibfield
  {journal} {\bibinfo  {journal} {Prog. Part. Nucl. Phys.}\ }\textbf {\bibinfo
  {volume} {63}},\ \bibinfo {pages} {117 } (\bibinfo {year}
  {2009})}\BibitemShut {NoStop}%
\bibitem [{\citenamefont {Drut}\ \emph {et~al.}(2011)\citenamefont {Drut},
  \citenamefont {L\"ahde},\ and\ \citenamefont {Ten}}]{Drut2011}%
  \BibitemOpen
  \bibfield  {author} {\bibinfo {author} {\bibfnamefont {J.~E.}\ \bibnamefont
  {Drut}}, \bibinfo {author} {\bibfnamefont {T.~A.}\ \bibnamefont {L\"ahde}},\
  and\ \bibinfo {author} {\bibfnamefont {T.}~\bibnamefont {Ten}},\ }\bibfield
  {title} {\bibinfo {title} {Momentum distribution and contact of the unitary
  fermi gas},\ }\href {https://link.aps.org/doi/10.1103/PhysRevLett.106.205302}
  {\bibfield  {journal} {\bibinfo  {journal} {Phys. Rev. Lett.}\ }\textbf
  {\bibinfo {volume} {106}},\ \bibinfo {pages} {205302} (\bibinfo {year}
  {2011})}\BibitemShut {NoStop}%
\bibitem [{\citenamefont {Liu}(2013)}]{Liu2013}%
  \BibitemOpen
  \bibfield  {author} {\bibinfo {author} {\bibfnamefont {X.-J.}\ \bibnamefont
  {Liu}},\ }\bibfield  {title} {\bibinfo {title} {Virial expansion for a
  strongly correlated fermi system and its application to ultracold atomic
  fermi gases},\ }\href {https://doi.org/10.1016/j.physrep.2012.10.004}
  {\bibfield  {journal} {\bibinfo  {journal} {Physics Reports}\ }\textbf
  {\bibinfo {volume} {524}},\ \bibinfo {pages} {37} (\bibinfo {year}
  {2013})}\BibitemShut {NoStop}%
\bibitem [{\citenamefont {Beth}\ and\ \citenamefont
  {Uhlenbeck}(1937)}]{Beth1937}%
  \BibitemOpen
  \bibfield  {author} {\bibinfo {author} {\bibfnamefont {E.}~\bibnamefont
  {Beth}}\ and\ \bibinfo {author} {\bibfnamefont {G.~E.}\ \bibnamefont
  {Uhlenbeck}},\ }\bibfield  {title} {\bibinfo {title} {The quantum theory of
  the non-ideal gas. ii. behaviour at low temperatures},\ }\href
  {https://doi.org/10.1016/S0031-8914(37)80189-5} {\bibfield  {journal}
  {\bibinfo  {journal} {Physica (Utrecht)}\ }\textbf {\bibinfo {volume} {4}},\
  \bibinfo {pages} {915} (\bibinfo {year} {1937})}\BibitemShut {NoStop}%
\bibitem [{\citenamefont {Lee}\ and\ \citenamefont
  {Sch\"afer}(2006)}]{Lee2006}%
  \BibitemOpen
  \bibfield  {author} {\bibinfo {author} {\bibfnamefont {D.}~\bibnamefont
  {Lee}}\ and\ \bibinfo {author} {\bibfnamefont {T.}~\bibnamefont
  {Sch\"afer}},\ }\bibfield  {title} {\bibinfo {title} {Cold dilute neutron
  matter on the lattice. i. lattice virial coefficients and large scattering
  lengths},\ }\href {https://doi.org/10.1103/PhysRevC.73.015201} {\bibfield
  {journal} {\bibinfo  {journal} {Phys. Rev. C}\ }\textbf {\bibinfo {volume}
  {73}},\ \bibinfo {pages} {015201} (\bibinfo {year} {2006})}\BibitemShut
  {NoStop}%
\bibitem [{\citenamefont {Liu}\ \emph {et~al.}(2009)\citenamefont {Liu},
  \citenamefont {Hu},\ and\ \citenamefont {Drummond}}]{Liu2009}%
  \BibitemOpen
  \bibfield  {author} {\bibinfo {author} {\bibfnamefont {X.-J.}\ \bibnamefont
  {Liu}}, \bibinfo {author} {\bibfnamefont {H.}~\bibnamefont {Hu}},\ and\
  \bibinfo {author} {\bibfnamefont {P.~D.}\ \bibnamefont {Drummond}},\
  }\bibfield  {title} {\bibinfo {title} {Virial expansion for a strongly
  correlated fermi gas},\ }\href
  {https://doi.org/10.1103/PhysRevLett.102.160401} {\bibfield  {journal}
  {\bibinfo  {journal} {Phys. Rev. Lett.}\ }\textbf {\bibinfo {volume} {102}},\
  \bibinfo {pages} {160401} (\bibinfo {year} {2009})}\BibitemShut {NoStop}%
\bibitem [{\citenamefont {Bedaque}\ and\ \citenamefont
  {Rupak}(2003)}]{Bedaque2003}%
  \BibitemOpen
  \bibfield  {author} {\bibinfo {author} {\bibfnamefont {P.~F.}\ \bibnamefont
  {Bedaque}}\ and\ \bibinfo {author} {\bibfnamefont {G.}~\bibnamefont
  {Rupak}},\ }\bibfield  {title} {\bibinfo {title} {Dilute resonating gases and
  the third virial coefficient},\ }\href
  {https://doi.org/10.1103/PhysRevB.67.174513} {\bibfield  {journal} {\bibinfo
  {journal} {Phys. Rev. B}\ }\textbf {\bibinfo {volume} {67}},\ \bibinfo
  {pages} {174513} (\bibinfo {year} {2003})}\BibitemShut {NoStop}%
\bibitem [{\citenamefont {Kaplan}\ and\ \citenamefont
  {Sun}(2011)}]{Kaplan2011}%
  \BibitemOpen
  \bibfield  {author} {\bibinfo {author} {\bibfnamefont {D.~B.}\ \bibnamefont
  {Kaplan}}\ and\ \bibinfo {author} {\bibfnamefont {S.}~\bibnamefont {Sun}},\
  }\bibfield  {title} {\bibinfo {title} {New field-theoretic method for the
  virial expansion},\ }\href {https://doi.org/10.1103/PhysRevLett.107.030601}
  {\bibfield  {journal} {\bibinfo  {journal} {Phys. Rev. Lett.}\ }\textbf
  {\bibinfo {volume} {107}},\ \bibinfo {pages} {030601} (\bibinfo {year}
  {2011})}\BibitemShut {NoStop}%
\bibitem [{\citenamefont {Leyronas}(2011)}]{Leyronas2011}%
  \BibitemOpen
  \bibfield  {author} {\bibinfo {author} {\bibfnamefont {X.}~\bibnamefont
  {Leyronas}},\ }\bibfield  {title} {\bibinfo {title} {Virial expansion with
  feynman diagrams},\ }\href {https://doi.org/10.1103/PhysRevA.84.053633}
  {\bibfield  {journal} {\bibinfo  {journal} {Phys. Rev. A}\ }\textbf {\bibinfo
  {volume} {84}},\ \bibinfo {pages} {053633} (\bibinfo {year}
  {2011})}\BibitemShut {NoStop}%
\bibitem [{\citenamefont {Castin}\ and\ \citenamefont
  {Werner}(2013)}]{Castin2013}%
  \BibitemOpen
  \bibfield  {author} {\bibinfo {author} {\bibfnamefont {Y.}~\bibnamefont
  {Castin}}\ and\ \bibinfo {author} {\bibfnamefont {F.}~\bibnamefont
  {Werner}},\ }\bibfield  {title} {\bibinfo {title} {Le troisième coefficient
  du viriel du gaz de bose unitaire},\ }\href
  {https://doi.org/10.1139/cjp-2012-0569} {\bibfield  {journal} {\bibinfo
  {journal} {Can. J. Phys.}\ }\textbf {\bibinfo {volume} {91}},\ \bibinfo
  {pages} {382} (\bibinfo {year} {2013})}\BibitemShut {NoStop}%
\bibitem [{\citenamefont {Gao}\ \emph {et~al.}(2015)\citenamefont {Gao},
  \citenamefont {Endo},\ and\ \citenamefont {Castin}}]{Gao2015}%
  \BibitemOpen
  \bibfield  {author} {\bibinfo {author} {\bibfnamefont {C.}~\bibnamefont
  {Gao}}, \bibinfo {author} {\bibfnamefont {S.}~\bibnamefont {Endo}},\ and\
  \bibinfo {author} {\bibfnamefont {Y.}~\bibnamefont {Castin}},\ }\bibfield
  {title} {\bibinfo {title} {The third virial coefficient of a two-component
  unitary fermi gas across an efimov-effect threshold},\ }\href
  {https://doi.org/10.1209/0295-5075/109/16003} {\bibfield  {journal} {\bibinfo
   {journal} {EPL}\ }\textbf {\bibinfo {volume} {109}},\ \bibinfo {pages}
  {16003} (\bibinfo {year} {2015})}\BibitemShut {NoStop}%
\bibitem [{\citenamefont {Rakshit}\ \emph {et~al.}(2012)\citenamefont
  {Rakshit}, \citenamefont {Daily},\ and\ \citenamefont {Blume}}]{Rakshit2012}%
  \BibitemOpen
  \bibfield  {author} {\bibinfo {author} {\bibfnamefont {D.}~\bibnamefont
  {Rakshit}}, \bibinfo {author} {\bibfnamefont {K.~M.}\ \bibnamefont {Daily}},\
  and\ \bibinfo {author} {\bibfnamefont {D.}~\bibnamefont {Blume}},\ }\bibfield
   {title} {\bibinfo {title} {Natural and unnatural parity states of small
  trapped equal-mass two-component fermi gases at unitarity and fourth-order
  virial coefficient},\ }\href {https://doi.org/10.1103/PhysRevA.85.033634}
  {\bibfield  {journal} {\bibinfo  {journal} {Phys. Rev. A}\ }\textbf {\bibinfo
  {volume} {85}},\ \bibinfo {pages} {033634} (\bibinfo {year}
  {2012})}\BibitemShut {NoStop}%
\bibitem [{\citenamefont {Endo}\ and\ \citenamefont {Castin}(2015)}]{Endo2015}%
  \BibitemOpen
  \bibfield  {author} {\bibinfo {author} {\bibfnamefont {S.}~\bibnamefont
  {Endo}}\ and\ \bibinfo {author} {\bibfnamefont {Y.}~\bibnamefont {Castin}},\
  }\bibfield  {title} {\bibinfo {title} {Absence of a four-body efimov effect
  in the $2+2$ fermionic problem},\ }\href
  {https://doi.org/10.1103/PhysRevA.92.053624} {\bibfield  {journal} {\bibinfo
  {journal} {Phys. Rev. A}\ }\textbf {\bibinfo {volume} {92}},\ \bibinfo
  {pages} {053624} (\bibinfo {year} {2015})}\BibitemShut {NoStop}%
\bibitem [{\citenamefont {Yan}\ and\ \citenamefont {Blume}(2016)}]{Yan2016}%
  \BibitemOpen
  \bibfield  {author} {\bibinfo {author} {\bibfnamefont {Y.}~\bibnamefont
  {Yan}}\ and\ \bibinfo {author} {\bibfnamefont {D.}~\bibnamefont {Blume}},\
  }\bibfield  {title} {\bibinfo {title} {Path-integral monte carlo
  determination of the fourth-order virial coefficient for a unitary
  two-component fermi gas with zero-range interactions},\ }\href
  {https://doi.org/10.1103/PhysRevLett.116.230401} {\bibfield  {journal}
  {\bibinfo  {journal} {Phys. Rev. Lett.}\ }\textbf {\bibinfo {volume} {116}},\
  \bibinfo {pages} {230401} (\bibinfo {year} {2016})}\BibitemShut {NoStop}%
\bibitem [{\citenamefont {Endo}\ and\ \citenamefont {Castin}(2016)}]{Endo2016}%
  \BibitemOpen
  \bibfield  {author} {\bibinfo {author} {\bibfnamefont {S.}~\bibnamefont
  {Endo}}\ and\ \bibinfo {author} {\bibfnamefont {Y.}~\bibnamefont {Castin}},\
  }\bibfield  {title} {\bibinfo {title} {The interaction-sensitive states of a
  trapped two-component ideal fermi gas and application to the virial expansion
  of the unitary fermi gas},\ }\href
  {https://doi.org/10.1088/1751-8113/49/26/265301} {\bibfield  {journal}
  {\bibinfo  {journal} {J. Phys. A}\ }\textbf {\bibinfo {volume} {49}},\
  \bibinfo {pages} {265301} (\bibinfo {year} {2016})}\BibitemShut {NoStop}%
\bibitem [{\citenamefont {Ngampruetikorn}\ \emph {et~al.}(2015)\citenamefont
  {Ngampruetikorn}, \citenamefont {Parish},\ and\ \citenamefont
  {Levinsen}}]{Ngampruetikorn2015}%
  \BibitemOpen
  \bibfield  {author} {\bibinfo {author} {\bibfnamefont {V.}~\bibnamefont
  {Ngampruetikorn}}, \bibinfo {author} {\bibfnamefont {M.~M.}\ \bibnamefont
  {Parish}},\ and\ \bibinfo {author} {\bibfnamefont {J.}~\bibnamefont
  {Levinsen}},\ }\bibfield  {title} {\bibinfo {title} {High-temperature limit
  of the resonant fermi gas},\ }\href
  {https://doi.org/10.1103/PhysRevA.91.013606} {\bibfield  {journal} {\bibinfo
  {journal} {Phys. Rev. A}\ }\textbf {\bibinfo {volume} {91}},\ \bibinfo
  {pages} {013606} (\bibinfo {year} {2015})}\BibitemShut {NoStop}%
\bibitem [{\citenamefont {Liu}\ and\ \citenamefont {Hu}(2010)}]{Liu2010}%
  \BibitemOpen
  \bibfield  {author} {\bibinfo {author} {\bibfnamefont {X.-J.}\ \bibnamefont
  {Liu}}\ and\ \bibinfo {author} {\bibfnamefont {H.}~\bibnamefont {Hu}},\
  }\bibfield  {title} {\bibinfo {title} {Virial expansion for a strongly
  correlated fermi gas with imbalanced spin populations},\ }\href
  {https://doi.org/10.1103/PhysRevA.82.043626} {\bibfield  {journal} {\bibinfo
  {journal} {Phys. Rev. A}\ }\textbf {\bibinfo {volume} {82}},\ \bibinfo
  {pages} {043626} (\bibinfo {year} {2010})}\BibitemShut {NoStop}%
\bibitem [{\citenamefont {Endo}(2020)}]{Endo2020}%
  \BibitemOpen
  \bibfield  {author} {\bibinfo {author} {\bibfnamefont {S.}~\bibnamefont
  {Endo}},\ }\bibfield  {title} {\bibinfo {title} {Virial expansion
  coefficients in the unitary fermi gas},\ }\href
  {https://doi.org/10.21468/SciPostPhysProc.3.049} {\bibfield  {journal}
  {\bibinfo  {journal} {SciPost Phys. Proc.}\ }\textbf {\bibinfo {volume}
  {3}},\ \bibinfo {pages} {49} (\bibinfo {year} {2020})}\BibitemShut {NoStop}%
\bibitem [{\citenamefont {Enss}(2012)}]{Enss2012}%
  \BibitemOpen
  \bibfield  {author} {\bibinfo {author} {\bibfnamefont {T.}~\bibnamefont
  {Enss}},\ }\bibfield  {title} {\bibinfo {title} {Quantum critical transport
  in the unitary fermi gas},\ }\href
  {https://link.aps.org/doi/10.1103/PhysRevA.86.013616} {\bibfield  {journal}
  {\bibinfo  {journal} {Phys. Rev. A}\ }\textbf {\bibinfo {volume} {86}},\
  \bibinfo {pages} {013616} (\bibinfo {year} {2012})}\BibitemShut {NoStop}%
\bibitem [{\citenamefont {Sanner}\ \emph {et~al.}(2011)\citenamefont {Sanner},
  \citenamefont {Su}, \citenamefont {Keshet}, \citenamefont {Huang},
  \citenamefont {Gillen}, \citenamefont {Gommers},\ and\ \citenamefont
  {Ketterle}}]{Sanner2011}%
  \BibitemOpen
  \bibfield  {author} {\bibinfo {author} {\bibfnamefont {C.}~\bibnamefont
  {Sanner}}, \bibinfo {author} {\bibfnamefont {E.~J.}\ \bibnamefont {Su}},
  \bibinfo {author} {\bibfnamefont {A.}~\bibnamefont {Keshet}}, \bibinfo
  {author} {\bibfnamefont {W.}~\bibnamefont {Huang}}, \bibinfo {author}
  {\bibfnamefont {J.}~\bibnamefont {Gillen}}, \bibinfo {author} {\bibfnamefont
  {R.}~\bibnamefont {Gommers}},\ and\ \bibinfo {author} {\bibfnamefont
  {W.}~\bibnamefont {Ketterle}},\ }\bibfield  {title} {\bibinfo {title}
  {Speckle imaging of spin fluctuations in a strongly interacting fermi gas},\
  }\href {https://doi.org/10.1103/PhysRevLett.106.010402} {\bibfield  {journal}
  {\bibinfo  {journal} {Phys. Rev. Lett.}\ }\textbf {\bibinfo {volume} {106}},\
  \bibinfo {pages} {010402} (\bibinfo {year} {2011})}\BibitemShut {NoStop}%
\bibitem [{\citenamefont {Parish}\ \emph {et~al.}(2007)\citenamefont {Parish},
  \citenamefont {Marchetti}, \citenamefont {Lamacraft},\ and\ \citenamefont
  {Simons}}]{Parish2007}%
  \BibitemOpen
  \bibfield  {author} {\bibinfo {author} {\bibfnamefont {M.~M.}\ \bibnamefont
  {Parish}}, \bibinfo {author} {\bibfnamefont {F.~M.}\ \bibnamefont
  {Marchetti}}, \bibinfo {author} {\bibfnamefont {A.}~\bibnamefont
  {Lamacraft}},\ and\ \bibinfo {author} {\bibfnamefont {B.~D.}\ \bibnamefont
  {Simons}},\ }\bibfield  {title} {\bibinfo {title} {Finite-temperature phase
  diagram of a polarized fermi condensate},\ }\href
  {https://doi.org/10.1038/nphys520} {\bibfield  {journal} {\bibinfo  {journal}
  {Nat. Phys.}\ }\textbf {\bibinfo {volume} {3}},\ \bibinfo {pages} {124}
  (\bibinfo {year} {2007})}\BibitemShut {NoStop}%
\bibitem [{\citenamefont {Radzihovsky}\ and\ \citenamefont
  {Sheehy}(2010)}]{Radzihovsky2010}%
  \BibitemOpen
  \bibfield  {author} {\bibinfo {author} {\bibfnamefont {L.}~\bibnamefont
  {Radzihovsky}}\ and\ \bibinfo {author} {\bibfnamefont {D.~E.}\ \bibnamefont
  {Sheehy}},\ }\bibfield  {title} {\bibinfo {title} {Imbalanced
  feshbach-resonant fermi gases},\ }\href
  {https://doi.org/10.1088%2F0034-4885%2F73%2F7%2F076501} {\bibfield  {journal}
  {\bibinfo  {journal} {Rep. Prog. Phys.}\ }\textbf {\bibinfo {volume} {73}},\
  \bibinfo {pages} {076501} (\bibinfo {year} {2010})}\BibitemShut {NoStop}%
\bibitem [{\citenamefont {Chevy}\ and\ \citenamefont {Mora}(2010)}]{Chevy2010}%
  \BibitemOpen
  \bibfield  {author} {\bibinfo {author} {\bibfnamefont {F.}~\bibnamefont
  {Chevy}}\ and\ \bibinfo {author} {\bibfnamefont {C.}~\bibnamefont {Mora}},\
  }\bibfield  {title} {\bibinfo {title} {Ultra-cold polarized fermi gases},\
  }\href {https://doi.org/10.1088%2F0034-4885%2F73%2F11%2F112401} {\bibfield
  {journal} {\bibinfo  {journal} {Rep. Prog. Phys.}\ }\textbf {\bibinfo
  {volume} {73}},\ \bibinfo {pages} {112401} (\bibinfo {year}
  {2010})}\BibitemShut {NoStop}%
\bibitem [{\citenamefont {Gubbels}\ and\ \citenamefont
  {Stoof}(2013)}]{Gubbels2013}%
  \BibitemOpen
  \bibfield  {author} {\bibinfo {author} {\bibfnamefont {K.}~\bibnamefont
  {Gubbels}}\ and\ \bibinfo {author} {\bibfnamefont {H.}~\bibnamefont
  {Stoof}},\ }\bibfield  {title} {\bibinfo {title} {Imbalanced fermi gases at
  unitarity},\ }\href
  {http://www.sciencedirect.com/science/article/pii/S0370157312004115}
  {\bibfield  {journal} {\bibinfo  {journal} {Phys. Rep.}\ }\textbf {\bibinfo
  {volume} {525}},\ \bibinfo {pages} {255 } (\bibinfo {year}
  {2013})}\BibitemShut {NoStop}%
\bibitem [{\citenamefont {Birse}\ \emph {et~al.}(2005)\citenamefont {Birse},
  \citenamefont {Krippa}, \citenamefont {McGovern},\ and\ \citenamefont
  {Walet}}]{Birse2005}%
  \BibitemOpen
  \bibfield  {author} {\bibinfo {author} {\bibfnamefont {M.~C.}\ \bibnamefont
  {Birse}}, \bibinfo {author} {\bibfnamefont {B.}~\bibnamefont {Krippa}},
  \bibinfo {author} {\bibfnamefont {J.~A.}\ \bibnamefont {McGovern}},\ and\
  \bibinfo {author} {\bibfnamefont {N.~R.}\ \bibnamefont {Walet}},\ }\bibfield
  {title} {\bibinfo {title} {Pairing in many-fermion systems: an exact
  renormalisation group treatment},\ }\href
  {https://doi.org/https://doi.org/10.1016/j.physletb.2004.11.044} {\bibfield
  {journal} {\bibinfo  {journal} {Phys. Lett. B}\ }\textbf {\bibinfo {volume}
  {605}},\ \bibinfo {pages} {287 } (\bibinfo {year} {2005})}\BibitemShut
  {NoStop}%
\bibitem [{\citenamefont {Diehl}\ \emph {et~al.}(2007)\citenamefont {Diehl},
  \citenamefont {Gies}, \citenamefont {Pawlowski},\ and\ \citenamefont
  {Wetterich}}]{Diehl2007}%
  \BibitemOpen
  \bibfield  {author} {\bibinfo {author} {\bibfnamefont {S.}~\bibnamefont
  {Diehl}}, \bibinfo {author} {\bibfnamefont {H.}~\bibnamefont {Gies}},
  \bibinfo {author} {\bibfnamefont {J.~M.}\ \bibnamefont {Pawlowski}},\ and\
  \bibinfo {author} {\bibfnamefont {C.}~\bibnamefont {Wetterich}},\ }\bibfield
  {title} {\bibinfo {title} {Flow equations for the bcs-bec crossover},\ }\href
  {https://doi.org/10.1103/PhysRevA.76.021602} {\bibfield  {journal} {\bibinfo
  {journal} {Phys. Rev. A}\ }\textbf {\bibinfo {volume} {76}},\ \bibinfo
  {pages} {021602} (\bibinfo {year} {2007})}\BibitemShut {NoStop}%
\bibitem [{\citenamefont {Diehl}\ \emph {et~al.}(2010)\citenamefont {Diehl},
  \citenamefont {Floerchinger}, \citenamefont {Gies}, \citenamefont
  {Pawlowkski},\ and\ \citenamefont {Wetterich}}]{Diehl2009}%
  \BibitemOpen
  \bibfield  {author} {\bibinfo {author} {\bibfnamefont {S.}~\bibnamefont
  {Diehl}}, \bibinfo {author} {\bibfnamefont {S.}~\bibnamefont {Floerchinger}},
  \bibinfo {author} {\bibfnamefont {H.}~\bibnamefont {Gies}}, \bibinfo {author}
  {\bibfnamefont {J.}~\bibnamefont {Pawlowkski}},\ and\ \bibinfo {author}
  {\bibfnamefont {C.}~\bibnamefont {Wetterich}},\ }\bibfield  {title} {\bibinfo
  {title} {Functional renormalization group approach to the bcs-bec
  crossover},\ }\href {https://doi.org/https://doi.org/10.1002/andp.201010458}
  {\bibfield  {journal} {\bibinfo  {journal} {Ann. Phys. (Berl.)}\ }\textbf
  {\bibinfo {volume} {522}},\ \bibinfo {pages} {615} (\bibinfo {year}
  {2010})}\BibitemShut {NoStop}%
\bibitem [{\citenamefont {Scherer}\ \emph {et~al.}(2011)\citenamefont
  {Scherer}, \citenamefont {Floerchinger},\ and\ \citenamefont
  {Gies}}]{Scherer2010}%
  \BibitemOpen
  \bibfield  {author} {\bibinfo {author} {\bibfnamefont {M.~M.}\ \bibnamefont
  {Scherer}}, \bibinfo {author} {\bibfnamefont {S.}~\bibnamefont
  {Floerchinger}},\ and\ \bibinfo {author} {\bibfnamefont {H.}~\bibnamefont
  {Gies}},\ }\bibfield  {title} {\bibinfo {title} {Functional renormalization
  for the bardeen-cooper-schrieffer to bose-einstein condensation crossover},\
  }\href {https://doi.org/10.1098/rsta.2011.0072} {\bibfield  {journal}
  {\bibinfo  {journal} {Phil. Trans. R. Soc. A.}\ }\textbf {\bibinfo {volume}
  {369}},\ \bibinfo {pages} {2779} (\bibinfo {year} {2011})}\BibitemShut
  {NoStop}%
\bibitem [{\citenamefont {Boettcher}\ \emph {et~al.}(2012)\citenamefont
  {Boettcher}, \citenamefont {Pawlowski},\ and\ \citenamefont
  {Diehl}}]{Boettcher2012}%
  \BibitemOpen
  \bibfield  {author} {\bibinfo {author} {\bibfnamefont {I.}~\bibnamefont
  {Boettcher}}, \bibinfo {author} {\bibfnamefont {J.~M.}\ \bibnamefont
  {Pawlowski}},\ and\ \bibinfo {author} {\bibfnamefont {S.}~\bibnamefont
  {Diehl}},\ }\bibfield  {title} {\bibinfo {title} {Ultracold atoms and the
  functional renormalization group},\ }\href
  {https://doi.org/https://doi.org/10.1016/j.nuclphysbps.2012.06.004}
  {\bibfield  {journal} {\bibinfo  {journal} {Nucl. Phys. B. Proc. Supp.}\
  }\textbf {\bibinfo {volume} {228}},\ \bibinfo {pages} {63 } (\bibinfo {year}
  {2012})},\ \bibinfo {note} {“Physics at all scales: The Renormalization
  Group” Proceedings of the 49th Internationale Universitätswochen für
  Theoretische Physik}\BibitemShut {NoStop}%
\bibitem [{\citenamefont {Frank}\ \emph {et~al.}(2018)\citenamefont {Frank},
  \citenamefont {Lang},\ and\ \citenamefont {Zwerger}}]{Frank2018}%
  \BibitemOpen
  \bibfield  {author} {\bibinfo {author} {\bibfnamefont {B.}~\bibnamefont
  {Frank}}, \bibinfo {author} {\bibfnamefont {J.}~\bibnamefont {Lang}},\ and\
  \bibinfo {author} {\bibfnamefont {W.}~\bibnamefont {Zwerger}},\ }\bibfield
  {title} {\bibinfo {title} {Universal phase diagram and scaling functions of
  imbalanced fermi gases},\ }\href {https://doi.org/10.1134/S1063776118110031}
  {\bibfield  {journal} {\bibinfo  {journal} {JETP}\ }\textbf {\bibinfo
  {volume} {127}},\ \bibinfo {pages} {812} (\bibinfo {year}
  {2018})}\BibitemShut {NoStop}%
\bibitem [{MCN()}]{MCNote}%
  \BibitemOpen
  \href@noop {} {}\bibinfo {note} {No determinations of $T_c$ from Monte Carlo
  approaches in the balanced case are shown in the above summary. Several
  values have been computed before, which, if properly scaled to the continuum
  limit, show good agreement to the values shown
  here~\cite{Burovski2006,Goulko2010,Goulko2010b,Jensen2020}.}\BibitemShut
  {Stop}%
\bibitem [{\citenamefont {Zdybel}\ and\ \citenamefont
  {Jakubczyk}(2020)}]{Zdybel2020}%
  \BibitemOpen
  \bibfield  {author} {\bibinfo {author} {\bibfnamefont {P.}~\bibnamefont
  {Zdybel}}\ and\ \bibinfo {author} {\bibfnamefont {P.}~\bibnamefont
  {Jakubczyk}},\ }\bibfield  {title} {\bibinfo {title} {Quantum lifshitz points
  and fluctuation-induced first-order phase transitions in imbalanced fermi
  mixtures},\ }\href {https://doi.org/10.1103/PhysRevResearch.2.033486}
  {\bibfield  {journal} {\bibinfo  {journal} {Phys. Rev. Res.}\ }\textbf
  {\bibinfo {volume} {2}},\ \bibinfo {pages} {033486} (\bibinfo {year}
  {2020})}\BibitemShut {NoStop}%
\bibitem [{\citenamefont {Lobo}\ \emph {et~al.}(2006)\citenamefont {Lobo},
  \citenamefont {Recati}, \citenamefont {Giorgini},\ and\ \citenamefont
  {Stringari}}]{Lobo2006}%
  \BibitemOpen
  \bibfield  {author} {\bibinfo {author} {\bibfnamefont {C.}~\bibnamefont
  {Lobo}}, \bibinfo {author} {\bibfnamefont {A.}~\bibnamefont {Recati}},
  \bibinfo {author} {\bibfnamefont {S.}~\bibnamefont {Giorgini}},\ and\
  \bibinfo {author} {\bibfnamefont {S.}~\bibnamefont {Stringari}},\ }\bibfield
  {title} {\bibinfo {title} {Normal state of a polarized fermi gas at
  unitarity},\ }\href {https://link.aps.org/doi/10.1103/PhysRevLett.97.200403}
  {\bibfield  {journal} {\bibinfo  {journal} {Phys. Rev. Lett.}\ }\textbf
  {\bibinfo {volume} {97}},\ \bibinfo {pages} {200403} (\bibinfo {year}
  {2006})}\BibitemShut {NoStop}%
\bibitem [{\citenamefont {Nishida}\ and\ \citenamefont
  {Son}(2007{\natexlab{b}})}]{Nishida2007b}%
  \BibitemOpen
  \bibfield  {author} {\bibinfo {author} {\bibfnamefont {Y.}~\bibnamefont
  {Nishida}}\ and\ \bibinfo {author} {\bibfnamefont {D.~T.}\ \bibnamefont
  {Son}},\ }\bibfield  {title} {\bibinfo {title} {Fermi gas near unitarity
  around four and two spatial dimensions},\ }\href
  {https://link.aps.org/doi/10.1103/PhysRevA.75.063617} {\bibfield  {journal}
  {\bibinfo  {journal} {Phys. Rev. A}\ }\textbf {\bibinfo {volume} {75}},\
  \bibinfo {pages} {063617} (\bibinfo {year} {2007}{\natexlab{b}})}\BibitemShut
  {NoStop}%
\bibitem [{\citenamefont {Pini}\ \emph {et~al.}(2021)\citenamefont {Pini},
  \citenamefont {Pieri}, \citenamefont {Grimm},\ and\ \citenamefont
  {Strinati}}]{Pini2021}%
  \BibitemOpen
  \bibfield  {author} {\bibinfo {author} {\bibfnamefont {M.}~\bibnamefont
  {Pini}}, \bibinfo {author} {\bibfnamefont {P.}~\bibnamefont {Pieri}},
  \bibinfo {author} {\bibfnamefont {R.}~\bibnamefont {Grimm}},\ and\ \bibinfo
  {author} {\bibfnamefont {G.~C.}\ \bibnamefont {Strinati}},\ }\bibfield
  {title} {\bibinfo {title} {Beyond-mean-field description of a trapped unitary
  fermi gas with mass and population imbalance},\ }\href
  {https://doi.org/10.1103/PhysRevA.103.023314} {\bibfield  {journal} {\bibinfo
   {journal} {Phys. Rev. A}\ }\textbf {\bibinfo {volume} {103}},\ \bibinfo
  {pages} {023314} (\bibinfo {year} {2021})}\BibitemShut {NoStop}%
\bibitem [{\citenamefont {Goulko}\ and\ \citenamefont
  {Wingate}(2010)}]{Goulko2010}%
  \BibitemOpen
  \bibfield  {author} {\bibinfo {author} {\bibfnamefont {O.}~\bibnamefont
  {Goulko}}\ and\ \bibinfo {author} {\bibfnamefont {M.}~\bibnamefont
  {Wingate}},\ }\bibfield  {title} {\bibinfo {title} {Thermodynamics of
  balanced and slightly spin-imbalanced fermi gases at unitarity},\ }\href
  {https://link.aps.org/doi/10.1103/PhysRevA.82.053621} {\bibfield  {journal}
  {\bibinfo  {journal} {Phys. Rev. A}\ }\textbf {\bibinfo {volume} {82}},\
  \bibinfo {pages} {053621} (\bibinfo {year} {2010})}\BibitemShut {NoStop}%
\bibitem [{\citenamefont {Kashimura}\ \emph {et~al.}(2012)\citenamefont
  {Kashimura}, \citenamefont {Watanabe},\ and\ \citenamefont
  {Ohashi}}]{Kashimura2012}%
  \BibitemOpen
  \bibfield  {author} {\bibinfo {author} {\bibfnamefont {T.}~\bibnamefont
  {Kashimura}}, \bibinfo {author} {\bibfnamefont {R.}~\bibnamefont
  {Watanabe}},\ and\ \bibinfo {author} {\bibfnamefont {Y.}~\bibnamefont
  {Ohashi}},\ }\bibfield  {title} {\bibinfo {title} {Spin susceptibility and
  fluctuation corrections in the bcs-bec crossover regime of an ultracold fermi
  gas},\ }\href {https://link.aps.org/doi/10.1103/PhysRevA.86.043622}
  {\bibfield  {journal} {\bibinfo  {journal} {Phys. Rev. A}\ }\textbf {\bibinfo
  {volume} {86}},\ \bibinfo {pages} {043622} (\bibinfo {year}
  {2012})}\BibitemShut {NoStop}%
\bibitem [{\citenamefont {Mueller}(2011)}]{Mueller2011}%
  \BibitemOpen
  \bibfield  {author} {\bibinfo {author} {\bibfnamefont {E.~J.}\ \bibnamefont
  {Mueller}},\ }\bibfield  {title} {\bibinfo {title} {Evolution of the
  pseudogap in a polarized fermi gas},\ }\href
  {https://doi.org/10.1103/PhysRevA.83.053623} {\bibfield  {journal} {\bibinfo
  {journal} {Phys. Rev. A}\ }\textbf {\bibinfo {volume} {83}},\ \bibinfo
  {pages} {053623} (\bibinfo {year} {2011})}\BibitemShut {NoStop}%
\bibitem [{\citenamefont {Mukherjee}\ \emph {et~al.}(2017)\citenamefont
  {Mukherjee}, \citenamefont {Yan}, \citenamefont {Patel}, \citenamefont
  {Hadzibabic}, \citenamefont {Yefsah}, \citenamefont {Struck},\ and\
  \citenamefont {Zwierlein}}]{Mukherjee2016}%
  \BibitemOpen
  \bibfield  {author} {\bibinfo {author} {\bibfnamefont {B.}~\bibnamefont
  {Mukherjee}}, \bibinfo {author} {\bibfnamefont {Z.}~\bibnamefont {Yan}},
  \bibinfo {author} {\bibfnamefont {P.~B.}\ \bibnamefont {Patel}}, \bibinfo
  {author} {\bibfnamefont {Z.}~\bibnamefont {Hadzibabic}}, \bibinfo {author}
  {\bibfnamefont {T.}~\bibnamefont {Yefsah}}, \bibinfo {author} {\bibfnamefont
  {J.}~\bibnamefont {Struck}},\ and\ \bibinfo {author} {\bibfnamefont {M.~W.}\
  \bibnamefont {Zwierlein}},\ }\bibfield  {title} {\bibinfo {title}
  {Homogeneous atomic fermi gases},\ }\href
  {https://link.aps.org/doi/10.1103/PhysRevLett.118.123401} {\bibfield
  {journal} {\bibinfo  {journal} {Phys. Rev. Lett.}\ }\textbf {\bibinfo
  {volume} {118}},\ \bibinfo {pages} {123401} (\bibinfo {year}
  {2017})}\BibitemShut {NoStop}%
\bibitem [{\citenamefont {Hueck}\ \emph {et~al.}(2018)\citenamefont {Hueck},
  \citenamefont {Luick}, \citenamefont {Sobirey}, \citenamefont {Siegl},
  \citenamefont {Lompe},\ and\ \citenamefont {Moritz}}]{Hueck2017}%
  \BibitemOpen
  \bibfield  {author} {\bibinfo {author} {\bibfnamefont {K.}~\bibnamefont
  {Hueck}}, \bibinfo {author} {\bibfnamefont {N.}~\bibnamefont {Luick}},
  \bibinfo {author} {\bibfnamefont {L.}~\bibnamefont {Sobirey}}, \bibinfo
  {author} {\bibfnamefont {J.}~\bibnamefont {Siegl}}, \bibinfo {author}
  {\bibfnamefont {T.}~\bibnamefont {Lompe}},\ and\ \bibinfo {author}
  {\bibfnamefont {H.}~\bibnamefont {Moritz}},\ }\bibfield  {title} {\bibinfo
  {title} {Two-dimensional homogeneous fermi gases},\ }\href
  {https://link.aps.org/doi/10.1103/PhysRevLett.120.060402} {\bibfield
  {journal} {\bibinfo  {journal} {Phys. Rev. Lett.}\ }\textbf {\bibinfo
  {volume} {120}},\ \bibinfo {pages} {060402} (\bibinfo {year}
  {2018})}\BibitemShut {NoStop}%
\bibitem [{\citenamefont {Altman}\ \emph {et~al.}(2004)\citenamefont {Altman},
  \citenamefont {Demler},\ and\ \citenamefont {Lukin}}]{Altman2004}%
  \BibitemOpen
  \bibfield  {author} {\bibinfo {author} {\bibfnamefont {E.}~\bibnamefont
  {Altman}}, \bibinfo {author} {\bibfnamefont {E.}~\bibnamefont {Demler}},\
  and\ \bibinfo {author} {\bibfnamefont {M.~D.}\ \bibnamefont {Lukin}},\
  }\bibfield  {title} {\bibinfo {title} {Probing many-body states of ultracold
  atoms via noise correlations},\ }\href
  {https://link.aps.org/doi/10.1103/PhysRevA.70.013603} {\bibfield  {journal}
  {\bibinfo  {journal} {Phys. Rev. A}\ }\textbf {\bibinfo {volume} {70}},\
  \bibinfo {pages} {013603} (\bibinfo {year} {2004})}\BibitemShut {NoStop}%
\bibitem [{\citenamefont {Goulko}\ and\ \citenamefont
  {Wingate}(2011)}]{Goulko2010b}%
  \BibitemOpen
  \bibfield  {author} {\bibinfo {author} {\bibfnamefont {O.}~\bibnamefont
  {Goulko}}\ and\ \bibinfo {author} {\bibfnamefont {M.}~\bibnamefont
  {Wingate}},\ }\bibfield  {title} {\bibinfo {title} {{The imbalanced Fermi gas
  at unitarity}},\ }\href {https://doi.org/10.22323/1.105.0187} {\bibfield
  {journal} {\bibinfo  {journal} {PoS}\ }\textbf {\bibinfo {volume} {Lattice
  2010}},\ \bibinfo {pages} {187} (\bibinfo {year} {2011})}\BibitemShut
  {NoStop}%
\end{thebibliography}%
\end{document}